\documentclass[draft]{agujournal2019}
\usepackage{url} 
\usepackage[inline]{trackchanges} 
\usepackage{soul}
\usepackage{mathtools}
\usepackage{caption}
\usepackage{subcaption}
\usepackage{amsmath}
\usepackage{amsfonts}
\usepackage{amsthm}
\usepackage{amsbsy}
\usepackage{algorithm}
\usepackage{algpseudocode}
\usepackage{natbib}

\usepackage[textsize=tiny]{todonotes}

\definecolor{darkgreen}{rgb}{.15,.55,0}

\newcommand{\ym}[1]{\textcolor{black}{#1}}

\draftfalse

\journalname{}
\begin{document}
\title{Uncertainty Quantification of Fluid Leakage and Fault Instability in  Geologic CO$_{2}$ Storage}
\authors{Hannah Lu\affil{1,2}, Llu\'{i}s Sal\'{o}-Salgado\affil{1,3}, Youssef M. Marzouk\affil{2}, Ruben Juanes\affil{1,4}}
\affiliation{1}{Department of Civil and Environmental Engineering, Massachusetts Institute of Technology, Cambridge, MA, USA}
\affiliation{2}{Department of Aeronautics and Astronautics, Massachusetts Institute of Technology, Cambridge, MA, USA}
\affiliation{3}{Department of Earth and Planetary Sciences, Harvard University, Cambridge, MA, USA}
\affiliation{4}{Department of Earth, Atmospheric, and Planetary Sciences, Massachusetts Institute of Technology, Cambridge, MA, USA}
\correspondingauthor{Hannah Lu}{hannahlu@mit.edu}

\begin{keypoints}
\item We identify quantities of interest to assess coupled multiphase flow and geomechanics in anisotropic fault zones with uncertain parameters.
\item We use deep learning tools to build dynamical surrogates for the target quantities of interest.
\item Quantification of the uncertainty is accelerated by the accurate forecasting from deep-learning-based surrogates.
\end{keypoints}

\begin{abstract}
Geologic CO$_2$ storage is an important strategy for reducing greenhouse gas emissions to the atmosphere and mitigating climate change. In this process, coupling between mechanical deformation and fluid flow in fault zones is a key determinant of fault instability, induced seismicity, and CO$_2$ leakage. Using a recently developed methodology, PREDICT, we obtain probability distributions of the permeability tensor in faults from the stochastic placement of clay smears that accounts for geologic uncertainty. We build a comprehensive set of fault permeability scenarios from PREDICT and investigate the effects of uncertainties from the fault zone internal structure and composition on forecasts of CO$_2$ permanence and fault stability. To tackle the prohibitively expensive computational cost of the large number of simulations required to quantify uncertainty, we develop a deep-learning-based surrogate model capable of predicting flow migration, pressure buildup, and geomechanical responses in CO$_2$ storage operations. We also compare our probabilistic estimation of CO$_2$ leakage and fault instability with previous studies based on deterministic estimates of fault permeability. The results highlight the importance of including uncertainty and anisotropy in modeling of complex fault structures and improved management of geologic CO$_2$ storage projects.
\end{abstract}


\section{Introduction}
As a response to global climate change, carbon capture and storage (CCS) has been proposed as a mitigation technology to significantly reduce atmospheric CO$_2$ emissions and achieve net-zero emissions by 2050~\citep{metz2005ipcc,orr2009onshore,szulczewski2012lifetime,cozzi2020world,krevor2023subsurface}. Injecting CO$_2$ into geologic formations requires displacement or compression of the ambient groundwater and leads to pressure buildup in the storage aquifer. Potential hazards introduced from the operation include compromising the caprock by creating fractures and activating faults~\citep{birkholzer2009basin,zoback2012earthquake}, induced shear slip and triggered seismicity~\citep{rutqvist2007estimating,rutqvist2008coupled,chiaramonte2008seal,rutqvist2010coupled,morris2011large,morris2011study,cappa2011impact,cappa2011modeling,jha2014coupled, white2014geomechanical, white2016assessing,jagalur2018inferring}, CO$_2$ leakage and impacts on groundwater~\citep{keating2010impact,newell2018science,meguerdijian2021quantification}. Therefore, comprehensive uncertainty quantification and risk assessment are essential for safe reservoir engineering designs, proper mitigation plans, and long-term management of CO$_2$ storage.

Quantitative probabilistic risk assessment of CCS is highly challenging for two reasons. First, the modeling of CO$_2$ geological storage requires time consuming and computationally intensive simulations for coupling between multiphase flow and geomechanics~\citep{jha2014coupled,silva2024, krevor2023subsurface}. Computational challenges in multiscale modeling include various resolution requirements for different regions (e.g., highly resolved grids are needed around injection wells and fault zones) and extremely large spatial-temporal domains because CO$_2$ plume, pressure buildup and strain/stress responses propagate at different rates. Although various coupling schemes have been introduced to model the interactions between flow and geomechanics~\citep{dean2006comparison,jeannin2007accelerating,jha2007locally,mainguy2002coupling,minkoff2003coupled,settari1998coupled,settari2001advances,tran2004new,tran2005improved,kim2011stabilitya,kim2013rigorous,white2016block,both2017robust}, restrictive conditions may need to be satisfied for stability and long iterations may be needed for convergence of highly nonlinear systems. Second, the inherent uncertainty in highly heterogeneous porous media~\citep{kitanidis2015persistent,mallison2014unstructured}, together with the presence of faults and fractures~\citep{rinaldi2013modeling,morris2011study}, demands probabilistic descriptions of the material properties accounting for heterogeneity and/or anisotropy. Due to our incomplete knowledge of fault zones, stochastic modeling of the fault properties is imperative~\citep{salo2023fault}. Because parameterization of uncertainties in fault properties can be high-dimensional and complex, uncertainty quantification of CO$_2$ migration and fault stability requires running a large number of accurate numerical simulations, making the procedure prohibitively expensive.

In recent years, a number of deep-learning-based surrogate models have been developed as a promising alternative to expensive numerical simulators for subsurface problems. One approach (e.g.,~\citet{ZHU2018415,mo2019,TANG2020109456,WEN2021103223,WEN2021104009, WEN2023}) relies on data-driven learning of the underlying physics by approximating the mapping from diverse input field properties (e.g., permeability) to output state fields (e.g., fluid saturation and pressure buildup) \ym{given, as training data,} a large number of high-fidelity simulations. Therefore, the size of the neural network scales with the size of the grid and a convolutional neural network (CNN) architecture (e.g., U-Net~\citep{ronneberger2015u}) is used essentially to conduct image-to-image regression. A similar framework was later extended to coupled flow-geomechanics problems in~\citet{tang2022deep}. Alternatively, neural operators~\citep{lu2019deeponet,li2020fourier} are designed to find a discretization-invariant representation of the same mapping, so that the size of the network does not scale with grid resolution. Neural operators still require a large amount of data for training, however. More recently, graph neural network (GNN)-based surrogate models~\citep{sanchez2020learning,wu2022learning,ju2024learning} have been proposed to handle unstructured meshes with stencils that vary in size and shape, removing the constraint of having regular grids and greatly improving applicability to complex geological features such as faults and fractures. While significant computational gains can be achieved through cheap inference with the aforementioned deep-learning surrogate models, the computational costs of acquiring the training data and training the network can be substantial. A different approach, known as physics-informed neural networks (PINNs)~\citep{raissi2019physics}, integrates data with physics constraints in the form of a PDE-based loss function. PINNs have recently been applied to subsurface flow, transport and geomechanics problems~\citep{fuks2020limitations,he2020physics,haghighat2021physics,haghighat2022physics,amini2022physics,yan2022gradient}. The amount of training data needed for PINNs is less than in purely data-driven methods, but the neural network is usually more difficult to train because of the complex non-convex and multi-objective loss function~\citep{wang2020understanding,chen2018gradnorm}.

Constructing an efficient surrogate model to be used in uncertainty quantification and risk assessment for the coupled processes of mechanical deformation and fluid flow in fault zones is particularly challenging for several reasons. First, the complexity of the governing equations and the coupling of flow and geomechanics make PINNs particularly difficult and expensive to train. Second, purely data-driven methods require a substantial amount of training data, and in the data-limited regime, they are prone to overfitting. Additionally, the training data may not be sufficiently representative for the neural network to learn the underlying physics of state fields. For example, the uncertainties from the fault zone are very local and \ym{produce pressure buildup responses} only within the storage reservoir and the fault, providing much less variability in training data than a full domain permeability field. Lastly, fixed time-window methods are unsuitable for capturing the dynamic processes in CO2 sequestration, as these systems require models that can adapt to varying time horizons based on the occurrence of physical events such as fault slip. This underscores the need for a dynamic method, which will be explored in the next section.

Given the aforementioned challenges, we propose to use flow map learning (FML)~\citep{qin2019data,fu2020learning,qin2021deep,churchill2023flow}, a deep-learning-based numerical approximation for dynamical systems, to construct efficient surrogate models directly for the target quantities of interest (QoIs) in the coupled process of flow and geomechanics. The QoIs are low-dimensional quantities that represent the flow migration, pressure buildup and geomechanical responses in CO$_2$ storage operations, which can be used directly to monitor hazards like fault instability and induced seismicity. Compared with learning the full state field, learning the low-dimensional QoIs is computationally much more affordable, requires less training data and provides modeling outputs that are more directly relevant. The dynamical formulation of FML also allows for flexibility in the time horizon of interest. 

In section~\ref{sec:2}, we review the physics-based modeling of coupled flow and geomechanics, and specify the QoIs and sources of uncertainties in a representative two dimensional model. In section~\ref{sec:3}, we give an overview of the FML methodology with a detailed description of data preparation and training protocol. Finally, in section~\ref{sec:4}, we employ the FML surrogate models for accelerated uncertainty quantification to validate its accuracy, efficiency and robustness.

\section{Physics-based Modeling of Coupled Flow and Geomechanics}~\label{sec:2}
In this section, we first introduce the governing equations for coupled multiphase flow and geomechanics in a general form and then derive the quantities of interests related to fault instability in a specific example of CO$_2$ injection in a confined aquifer. At the end of the section, we introduce the sources of uncertainty from fault permeability and reservoir properties to the system and illustrate our methodology for modeling these uncertainties.

\subsection{Governing Equations}
We assume that the solid under consideration is in static equilibrium, the fluids are immiscible and the conditions are isothermal.  The governing equations for coupled multi-phase flow and geomechanics are obtained from conservation of mass and balance of linear momentum as follows:

\begin{equation}\label{eq:PDE}
\left\{
\begin{aligned}
&\frac{dm_\alpha}{dt} +\nabla_{\boldsymbol \xi} \cdot \mathbf w_\alpha = \rho_\alpha f_\alpha,\\
&\nabla_{\boldsymbol \xi}\cdot \boldsymbol \sigma +\rho_b\mathbf g = \mathbf 0,
\end{aligned}
\right.
\quad
\begin{aligned}
&\alpha = 1,\ldots, n_\text{phase},\\
&\boldsymbol \xi \in \Omega\setminus\partial \Omega,\quad  t\in [0,+\infty),
\end{aligned}
\end{equation}
where $\Omega$ is the spatial domain of interest, $\partial \Omega$ is its closed boundary, $m_\alpha$ denotes the mass of fluid phase $\alpha$, $\mathbf w_\alpha$ is the mass flux of fluid phase $\alpha$ relative to the solid skeleton, and $f_\alpha$ is the volumetric source term for phase $\alpha$, $\boldsymbol \sigma $ is the Cauchy total stress tensor, $\mathbf g$ is the gravity vector and $\rho_b$ is the bulk density. In an $n_\text{phase}$-phase fluid system, the bulk density can be expressed as
$
\rho_b = \phi\sum_{\beta}^{n_\text{phase}}\rho_\beta S_\beta +(1-\phi)\rho_s,
$
where $\rho_\beta$ and $S_\beta$ are the density and saturation of fluid phase $\beta$, $\rho_s$ is the density of the solid phase and $\phi$ is the true porosity (i.e., the ratio of the pore volume $V_p$ to the bulk volume $V_b$ in the current deformed configuration). The two-way coupling of the balance equation~\eqref{eq:PDE} is determined by virtue of poromechanics: changes in the pore fluid pressure lead to changes in \textit{effective stress}
and induce deformation of the porous medium while the deformation of the porous medium affects fluid mass content and fluid pressure. Following Biot's macroscopic theory of poroelasticity~\citep{biot1941general} and Coussy's incremental formulation of poromechanics for multiphase systems~\citep{coussy2004poromechanics}, we first decompose the total stress $\boldsymbol \sigma$ into two parts written in incremental form:
\begin{equation}
\delta \boldsymbol \sigma = \mathbf C_{dr} : \delta \boldsymbol \varepsilon -\sum_\beta^{n_\text{phase}} b_\beta\delta p_\beta \mathbf 1,
\end{equation}
where $\mathbf C_{dr}$ is the rank-$4$ drained elasticity tensor, $\boldsymbol \varepsilon$ is the strain tensor, $b_\beta$ are the Biot coefficients for individual phases such that $\sum_\beta b_\beta = b$ with $b$ being the Biot coefficient of the saturated porous material and $\mathbf 1$ is the rank-$2$ identity tensor. In the above formulation, the first term on the right-hand side represents the key concept of effective stress~\citep{bishop1959principal,bishop1963some}, which is responsible for deformation of the solid skeleton, and the second term is responsible for changes in the fluid pressures from all phases.

\subsection{Fault Instability from CO$_2$ Injection}\label{sec:risk}
We consider an example of CO$_2$ injection in a deep confined aquifer for the purpose of geologic carbon storage~\citep{cappa2011impact}. Figure~\ref{fig:domain} shows the initial conditions and the geometry of a two dimensional plane-strain model, where the domain of interest is $\Omega = [0,2000]$m $\times$ $[500, 2500]$m. The $100$m-thick storage aquifer is confined by two $150$m-thick layers of low-permeability caprocks. This multilayer system is disected by a pre-existing normal fault with a dip angle $\theta = 80^\circ$, an offset of $125$m and width of $2.5$m. The fault domain is denoted by $\Omega_f\subset \Omega$. Assuming a thermal gradient of $25^\circ$C/km and a hydrostatic gradient of $9.81$ MPa/km, the temperature and initial fluid pressure at several depths are listed in Table~\ref{table:1}. Given the listed temperature and pressure, CO$_2$ at supercritical condition is injected as a point source at $1500$m depth at a constant rate of $0.02$ kg/m/s. The boundaries are open for fluid flow with constant pressure and saturation except for the no-flow boundary condition at the left. Null displacement conditions are set normal to the left and bottom boundaries and stress is set to the right and top boundaries. Under normal faulting conditions, the vertical principal stress $\sigma_v$ due to gravity is larger than the horizontal principal stress $\sigma_h$ and a constant parameter $\lambda$ represents the initial ratio of the two (i.e., $\sigma_h = \lambda \sigma_v$). Table~\ref{table:2} shows the material properties of the reservoir, where $\phi_s, k_s,c_r$ and $\mathbf k_f$ are uncertain parameters characterizing our incomplete knowledge about the reservoir. Here we assume that the permeability of the storage aquifer is homogeneous and isotropic, i.e., $k_s$ is a scalar, and the permeability of the fault can be inhomogeneous and anisotropic, i.e., $\mathbf k_f(\boldsymbol \xi) = (\mathbf k_f^x(\boldsymbol \xi),\mathbf k_f^z(\boldsymbol \xi))$ for $\boldsymbol \xi\in \Omega_f$. The uncertainties arising from the fault permeability and the reservoir properties will be further discussed in section~\ref{sec: uncertainties}.

\vspace{0.4cm}

 \begin{minipage}{\textwidth}
  \begin{minipage}{0.4\textwidth}
    \centering
\includegraphics[height = 5 cm]{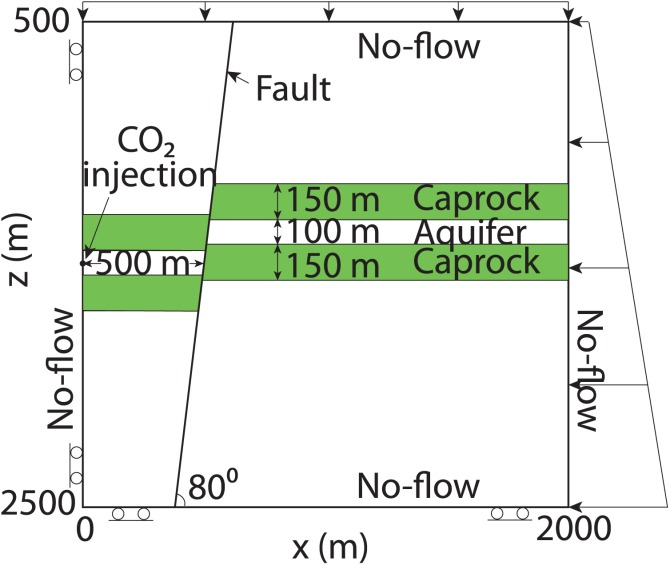}
    \captionof{figure}{Configuration of a 2D plane-strain model (from Cappa \&
Rutqvist, 2011a).}
    \label{fig:domain}
  \end{minipage}
\hfill
  \begin{minipage}{0.5\textwidth}
    \centering
    \vspace{1cm}
\begin{tabular}{  c c c  }
 \hline
& Temperature& Initial Pressure\\ 
 \hline
$0$m&$10$ [$^\circ$C]&$0.1$ [MPa]\\
 \hline
 $500$m & $22.5$ [$^\circ$C]&$5$ [MPa]\\ 
  \hline
  $1500$m&$47.5$ [$^\circ$C]&$14.72$ [MPa]\\
 \hline
 $2500$m & $72.5$ [$^\circ$C]&$24.63$ [MPa]\\ 
  \hline
\end{tabular}
\vspace{0.8cm}
      \captionof{table}{Temperature and initial fluid pressure at different depth.}
      \label{table:1}
    \end{minipage}
  \end{minipage}

\begin{table}[htp]
\begin{center}
\begin{tabular}{ l c c c c  } 
 \hline
Parameters & Storage Aquifer& Caprock&Other Aquifer&Fault\\ 
 \hline
Young's modulus&$10$ [GPa]&$10$ [GPa]&$10$ [GPa]&$5$ [GPa]\\
 \hline
 Poisson's ratio&$0.25$&$0.25$&$0.25$&$0.25$\\
  \hline
Porosity&$\phi_s$&$0.01$&$0.1$&$0.1$\\
 \hline
 Permeability&$k_s$ [m$^2$]&$10^{-19}$ [m$^2$]&$10^{-14}$ [m$^2$]&$\mathbf k_f$ [m$^2$]\\
 \hline
 Compressibility&$c_r$ [Pa$^{-1}$]&$c_r$ [Pa$^{-1}$]&$c_r$ [Pa$^{-1}$]&$c_r$ [Pa$^{-1}$]\\
 \hline
\end{tabular}
\end{center}
\caption{Material properties of the reservoir.}
\label{table:2}
\end{table}

\begin{figure}
\centering
\begin{subfigure}[b]{0.45\textwidth}
\centering
\includegraphics[width = \textwidth]{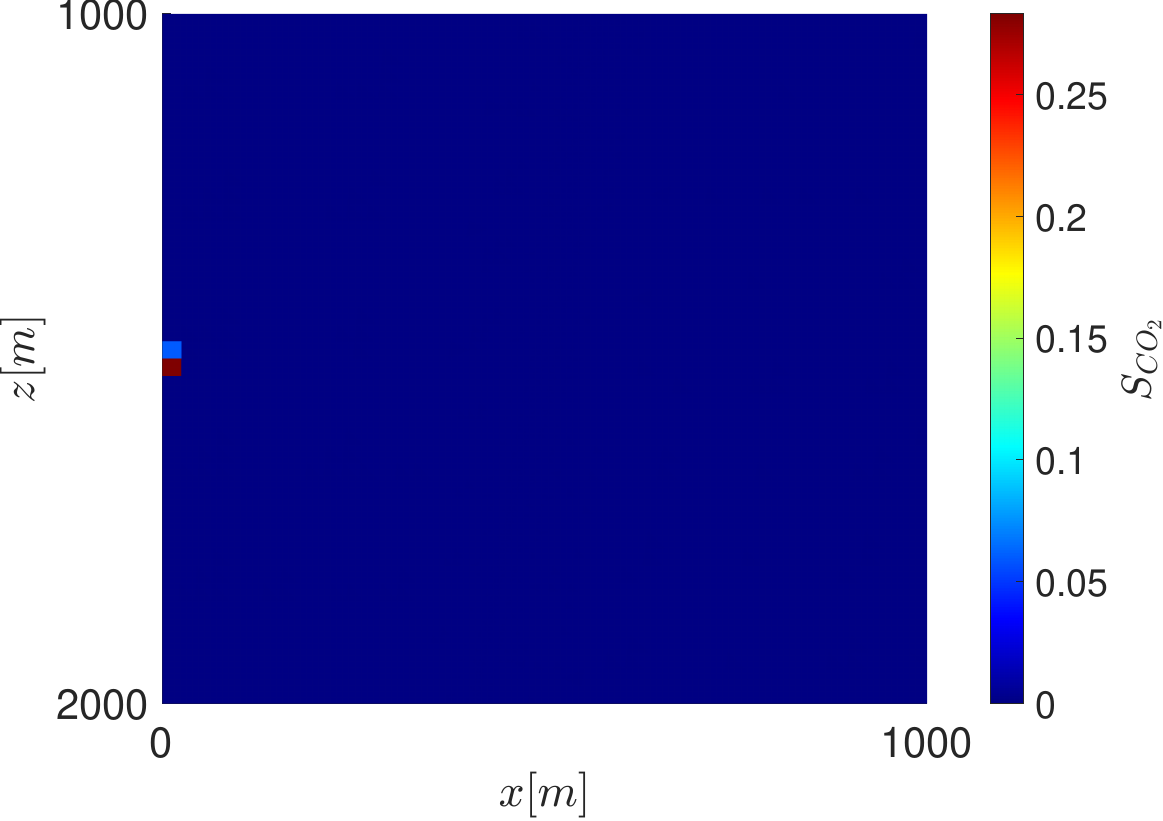}
\caption{CO$_2$ saturation at $t = 10$ days.}
\label{fig:2-a}
\end{subfigure}
\hfill
\begin{subfigure}[b]{0.45\textwidth}
\centering
\includegraphics[width = \textwidth]{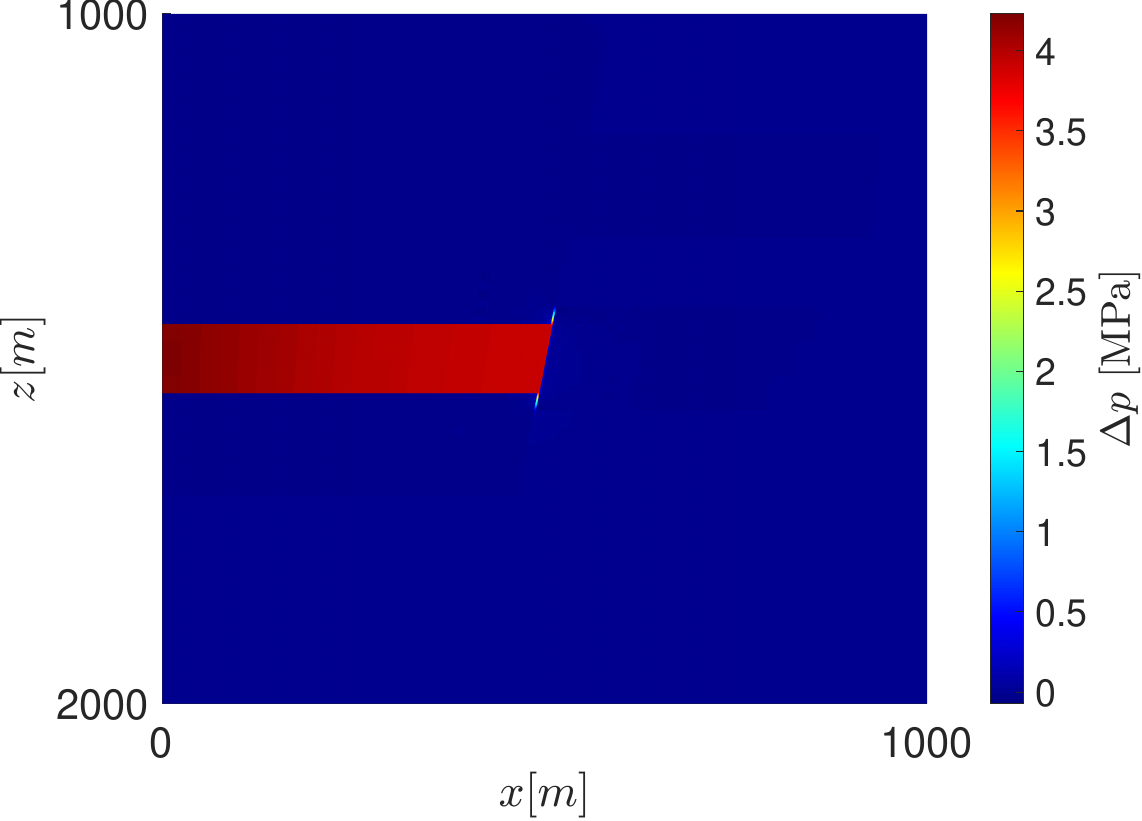}
\caption{Pressure buildup at $t = 10$ days.}
\label{fig:2-b}
\end{subfigure}

\begin{subfigure}[b]{0.45\textwidth}
\centering
\includegraphics[width = \textwidth]{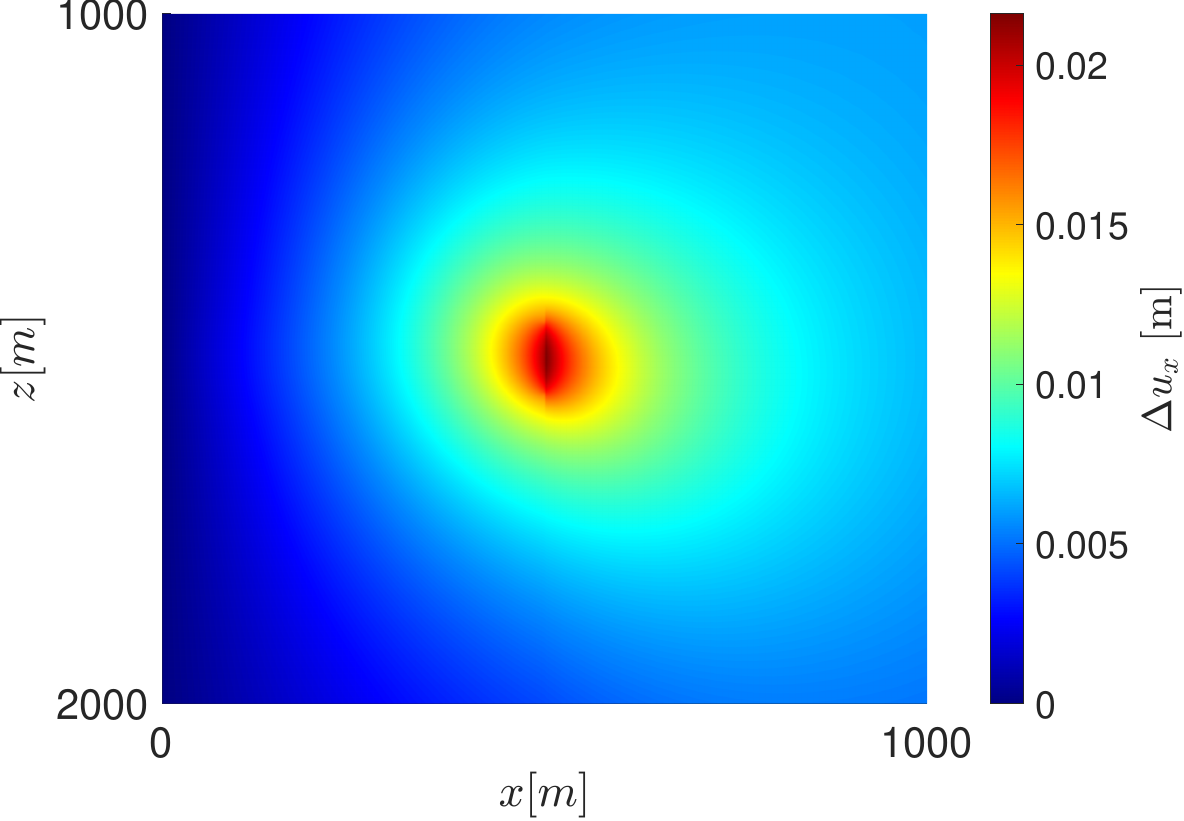}
\caption{Horizontal displacement field at $t = 10$ days.}
\label{fig:2-c}
\end{subfigure}
\hfill
\begin{subfigure}[b]{0.45\textwidth}
\centering
\includegraphics[width = \textwidth]{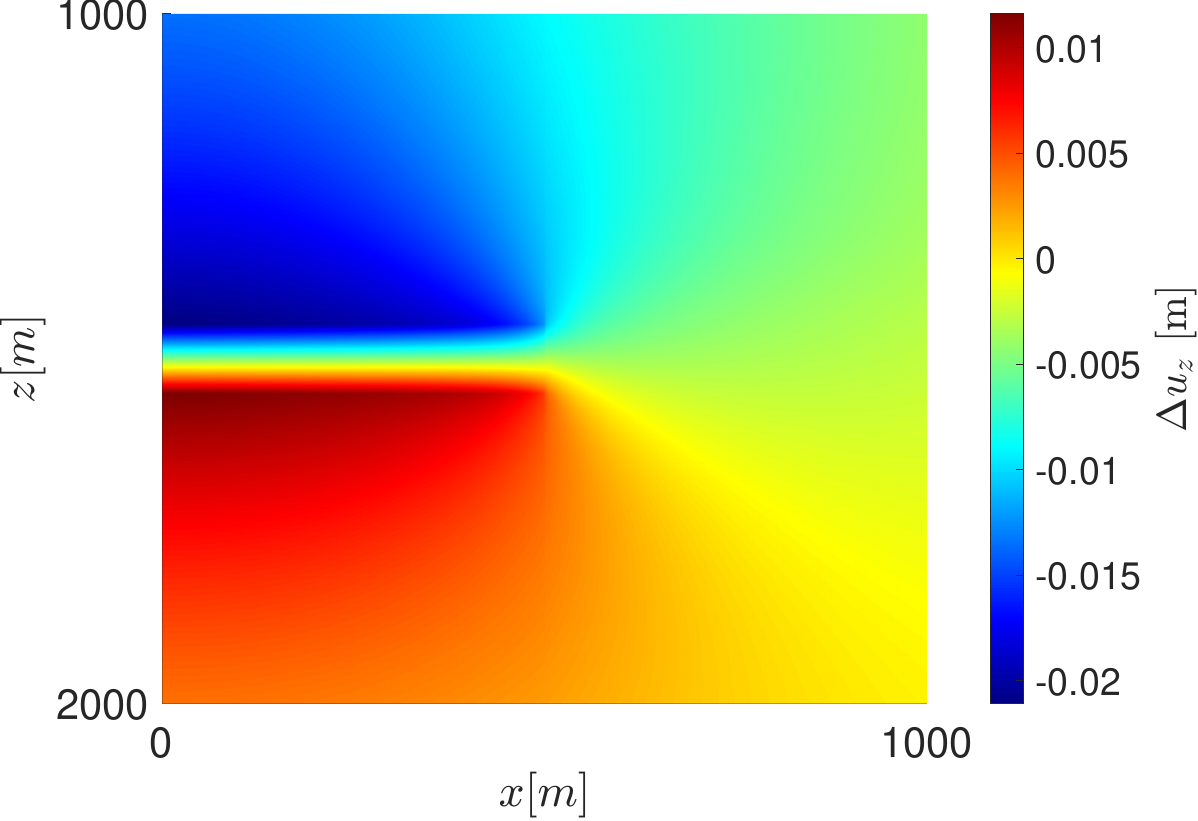}
\caption{Vertical displacement field at $t = 10$ days.}
\label{fig:2-d}
\end{subfigure}
\caption{A representative simulation with parameters: $\mathbf k_f = (10^{-16},10^{-16})$ [m$^2$], $\lambda = 0.7$, $k_s = 10^{-13}$ [m$^2$], $\phi_s = 0.1$ , $c_r = 1.45\times 10^{-10}$ [Pa$^{-1}$].}
\label{fig:solution}
\end{figure}

Figure~\ref{fig:solution} shows the results of a representative simulation with parameters $\mathbf k_f = (10^{-16},10^{-16})$ [m$^2$], $\lambda = 0.7$, $k_s = 10^{-13}$ [m$^2$], $\phi_s = 0.1$ , $c_r = 1.45\times 10^{-10}$ [Pa$^{-1}$]. After 10 days of injection, pressure builds up in the storage aquifer uniformly while CO$_2$ saturation migrates within a small distance from the injection well. The pressure buildup due to injection causes volumetric expansion of the aquifer (Figure~\ref{fig:2-c}-\ref{fig:2-d}). Additional details of the simulation framework are given in section~\ref{sec:MRST}.

\begin{figure}
\centering
\begin{subfigure}[b]{0.45\textwidth}
\centering
\includegraphics[width = \textwidth]{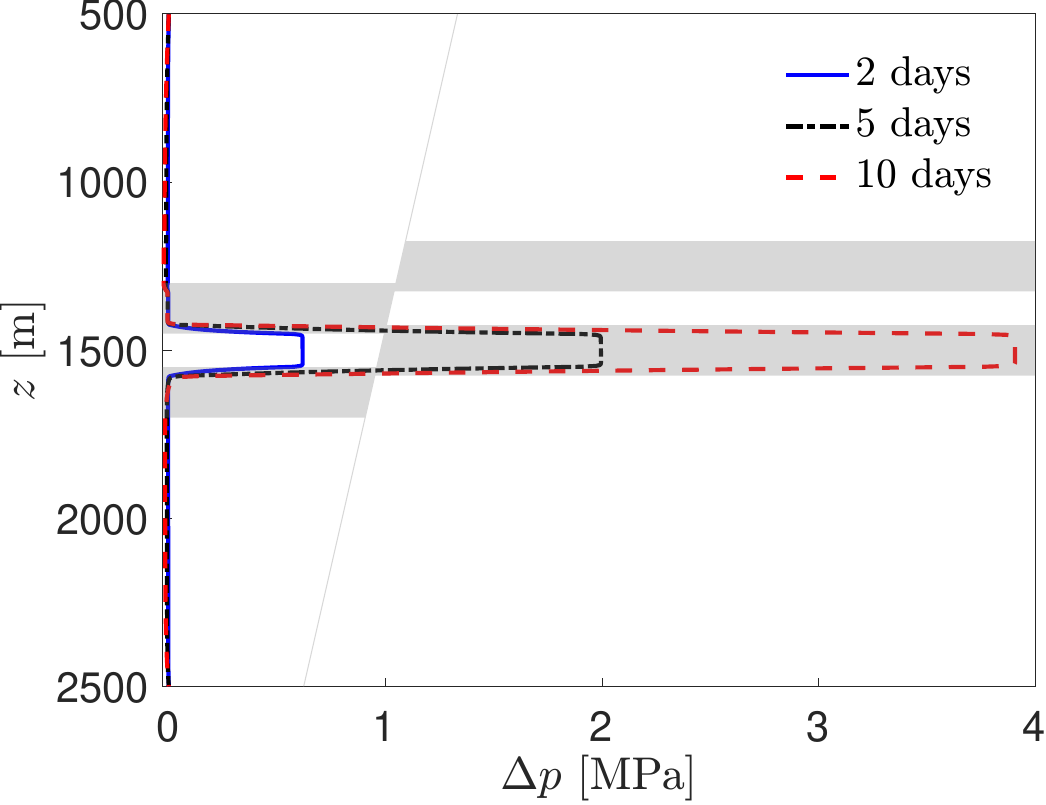}
\caption{Profiles of pressure buildup along the fault at $t=2, 5$ and $10$ days.}
\label{fig:3-a}
\end{subfigure}
\hfill
\begin{subfigure}[b]{0.45\textwidth}
\centering
\includegraphics[width = \textwidth]{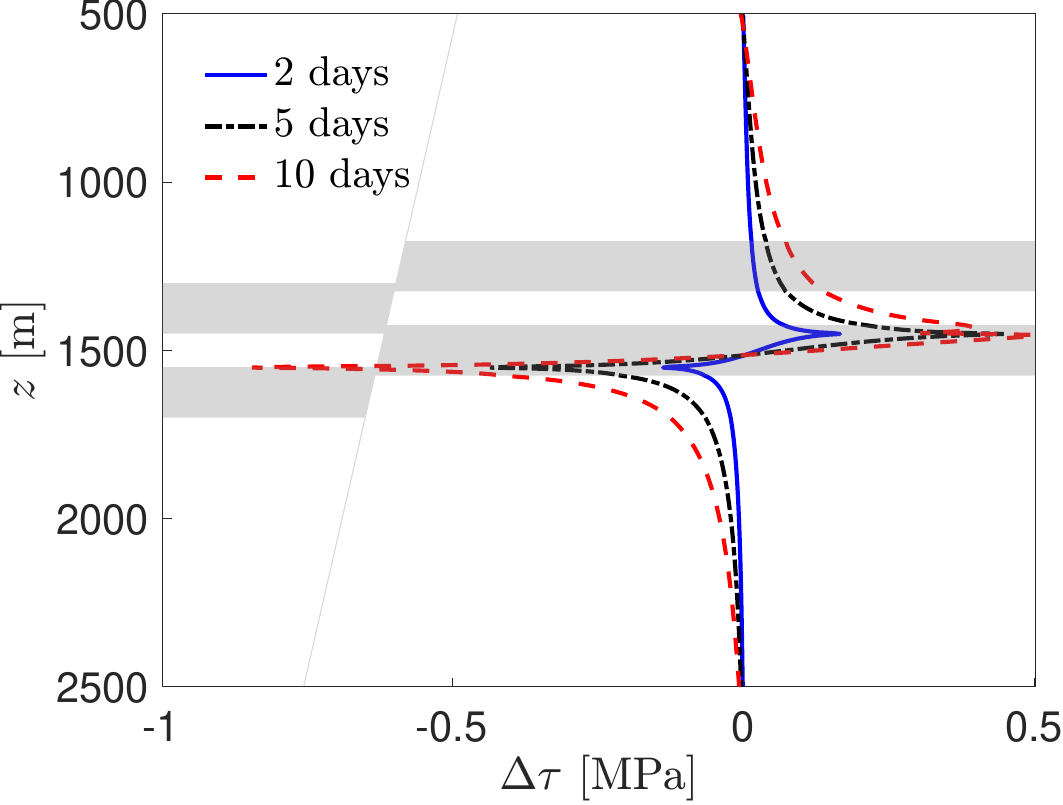}
\caption{Profiles of shear stress change along the fault at $t=2, 5$ and $10$ days.}
\label{fig:3-b}
\end{subfigure}

\begin{subfigure}[b]{0.45\textwidth}
\centering
\includegraphics[width = \textwidth]{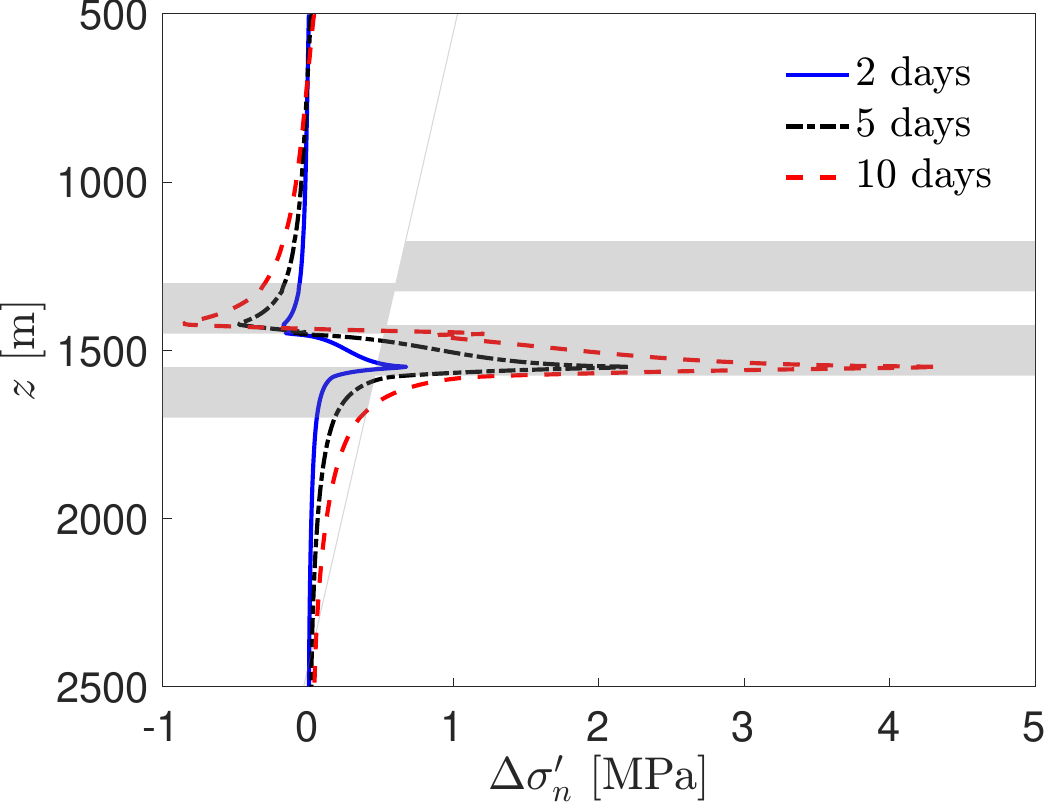}
\caption{Profiles of effective normal stress change along the fault at $t=2, 5$ and $10$ days.}
\label{fig:3-c}
\end{subfigure}
\hfill
\begin{subfigure}[b]{0.45\textwidth}
\centering
\includegraphics[width = \textwidth]{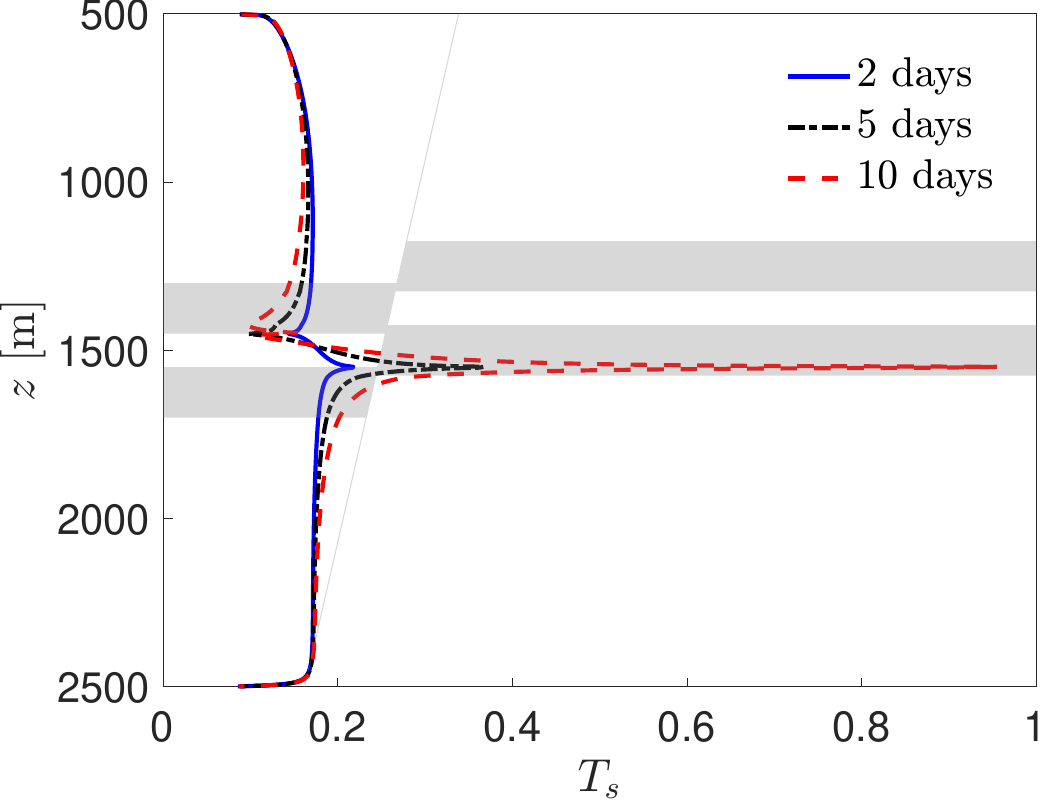}
\caption{Profiles of slip tendency along the fault at $t=2, 5$ and $10$ days.}
\label{fig:3-d}
\end{subfigure}
\caption{Vertical profiles of quantities related to  fault instability for a simulation, with parameters: $\mathbf k_f = (10^{-16},10^{-16})$ [m$^2$], $\lambda = 0.7$, $k_s = 10^{-13}$ [m$^2$], $\phi_s = 0.1$ , $c_r = 1.45\times 10^{-10}$ [Pa$^{-1}$].}
\label{fig:fault}
\end{figure}

A fault can undergo irreversible deformation when the shear stress applied to the fault plane becomes sufficiently high to exceed the shear strength of the fault, inducing shear displacement and/or fault reactivation. In order to estimate the fluid pressure threshold for fault reactivation, where the interconnected pore space is under an internal fluid pressure, the relevant quantity is the effective normal stress~\citep{terzaghi1923berechning}:
\begin{equation}\label{eq:effective-stress}
\sigma_n^\prime =  \sigma_n + p.
\end{equation}
Here, $\sigma_n^\prime$ is the effective normal stress, $\sigma_n$ is the total normal stress and $p$ is the fluid pressure (note that we use the convention that tensile normal stress is positive). Figures~\ref{fig:3-a}-\ref{fig:3-c} show that the volumetric expansion of the aquifer caused by pressure buildup due to injection results in an increase in the effective normal stress through the aquifer, and an increase in the magnitude of shear stress at the top and bottom boundaries of the aquifer, namely, at the depths of $1450$m and $1550$m. 

\begin{figure}
\centering
\includegraphics[width = 0.8\textwidth]{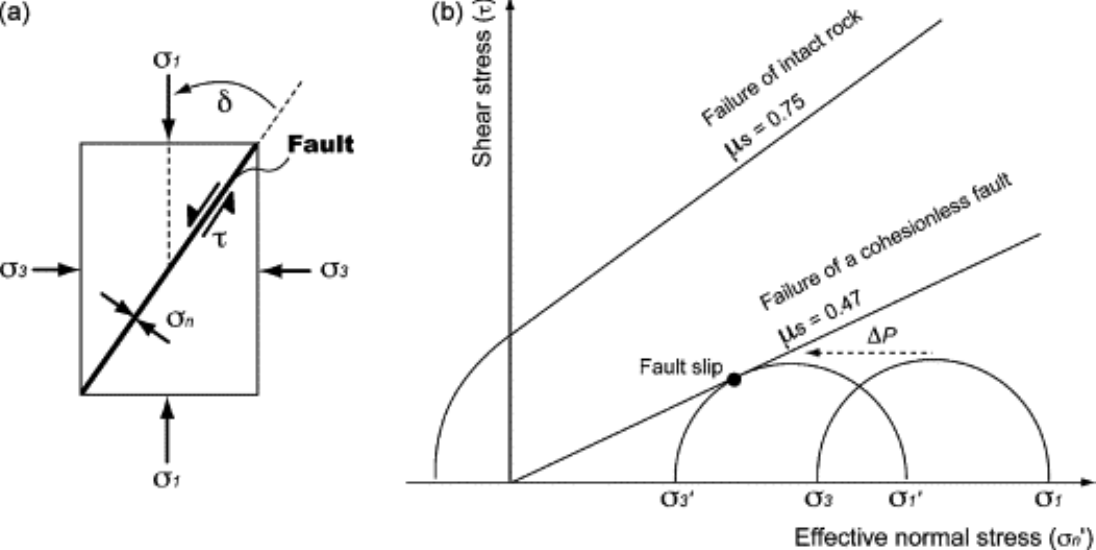}
\caption{(a). Normal and shear stresses resolved on a fault with a given orientation from the remote principal stresses; (b). Mohr diagram of shear stress ($\tau$) versus effective normal stress ($\sigma_n^\prime$) showing how the increasing fluid pressure ($\Delta p$) may activate a well-oriented cohesionless fault (fault slip) (from Cappa \& Rutqvist, 2011b).}
\label{fig:Coulomb-diagram}
\end{figure}

Considering hydromechanical interactions, failure analysis of a fault with a given dip angle reveals the fundamental relationship describing fault slip from the effective stress law~\eqref{eq:effective-stress} and the Coulomb failure criterion~\citep{coulomb1973essai}:
\begin{equation}\label{eq:shear-normal}
\tau = c+\mu_s\sigma_n^\prime,
\end{equation}
where $\tau$ is the critical shear stress for slip occurrence, $c$ is cohesion and $\mu_s$ is the static friction coefficient defined as $\mu_s = \tan(\psi)$, with $\psi$ being the friction angle. We assume $\psi = 25^\circ$ as in~\citet{cappa2011modeling}. The shear and normal stress acting on the fault plane can be computed from the two-dimensional principal stresses as
\begin{equation}\label{eq:shear-normal2}
\begin{aligned}
&\tau = \frac{\sigma_1-\sigma_3}{2}\sin 2\delta,\\
&\sigma_n = \frac{\sigma_1+\sigma_3}{2}-\frac{\sigma_1-\sigma_3}{2}\cos 2\delta,
\end{aligned}
\end{equation}
where $\sigma_1$ is the maximum principal stress, $\sigma_3$ is the minimum principal stress and $\delta$ is the angle between the fault plane and the $\sigma_1$ direction (Figure~\ref{fig:Coulomb-diagram}(a), taken from~\citet{cappa2011modeling}). In our case, $\sigma_1 = \sigma_h$, $\sigma_3 = \sigma_v$ and $\delta = 90^\circ-\theta$.

Equation~\eqref{eq:shear-normal} and Figure~\ref{fig:Coulomb-diagram}(b)~\citep{cappa2011modeling} indicate that increasing fluid pressure may induce shear slip along the fault. To estimate the potential for slip and reactivation of faults, we use the concept of slip tendency (or ambient stress ratio)~\citep{streit2004estimating}:
\begin{equation}\label{eq:slip-tendency}
T_{s,\text{max}} = \frac{\tau}{\sigma_n^\prime}.
\end{equation}
Here the slip tendency $T_{s,\text{max}}$ is the ratio of shear stress $\tau$ to effective normal stress $\sigma_n^\prime$ acting on the fault plane. Equation~\eqref{eq:shear-normal} also indicates that slip will be induced once the slip tendency $T_{s,\text{max}}$ exceeds the coefficient of static friction $\mu_s$. Figure~\ref{fig:3-d} shows the increase of the slip tendency profile and the high risk of fault instability at $t=10$ days at the bottom boundary of the aquifer ($1550$m).

\subsection{Uncertainties from Fault Permeability and Reservoir Properties}\label{sec: uncertainties}
Permeabilities of fault rock have been measured in a large variety of host rocks~\citep{faulkner2001can,wibberley2003internal,faulkner2006slip,lockner2009geometry}. The permeability of the fault core typically ranges from $10^{-17}$ to $10^{-21}$ m$^2$, while the areas surrounding the fault core appear to be intensively fractured with permeability ranging from $10^{-14}$ to $10^{-16}$ m$^2$. Besides the wide range of the fault permeability values, faults develop clay smears in normally consolidated, shallow (depth $<-3$km) siliciclastic sequences~\citep{vrolijk2016clay}. Therefore, an accurate modeling of fault permeability should account for heterogeneity, anisotropy and uncertainty. A new methodology, named PREDICT, was introduced recently in~\citet{salo2023fault} to compute probability distributions for the directional components (dip-normal and dip-parallel) of the fault permeability tensor from statistical samples for a set of geological variables. These variables, which include geometrical, compositional, and mechanical properties, allow multiple discretizations of the fault core to be populated with sand and clay smears, which can be used to upscale the permeability to a coarser scale. The flexibility of the user-defined upscaling can be used to assign fault permeability in subsequent reservoir-scale flow simulations.

\begin{figure}
\centering
\includegraphics[width = \textwidth]{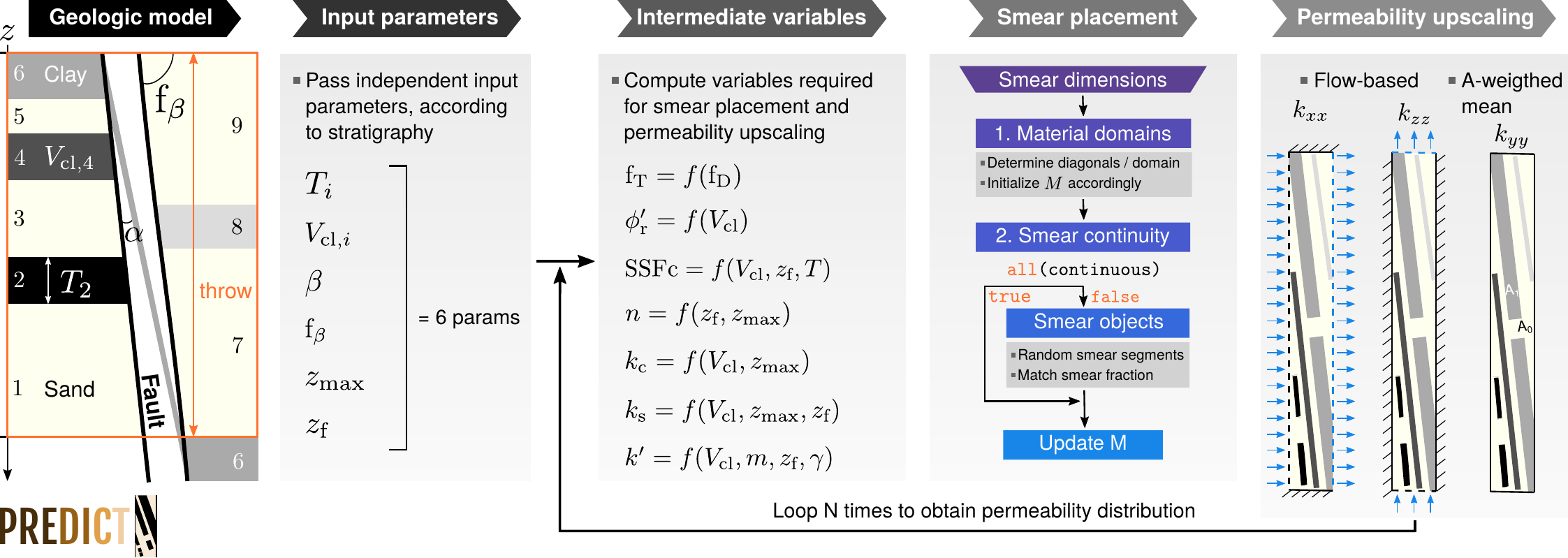}
\caption{PREDICT's workflow (left to right). Stratigraphic section is described by input parameters, which PREDICT uses to compute ranges and probability distributions for intermediate variables. For each fault section, fault thickness $f_T$, residual friction angle ($\phi_r^\prime$), critical shale smear factor (SSFc), porosity ($n$), permeability ($k$), and permeability anisotropy ratio ($k^\prime$) are sampled and used to place fault materials and assign their properties. Subscript “c” refers to clay smear, and “s” refers to sand smear. Permeability is upscaled using 2-D fault volume.}
\label{fig:PREDICT}
\end{figure}

A summary of PREDICT's workflow is shown in Figure~\ref{fig:PREDICT}. The computation is performed in a given throw window, in which PREDICT represents the fault core. The input parameters describing the faulted stratigraphy include the layer thickness ($T_{s,\text{max}}$), clay volume fraction ($V_{cl}$), dip angle ($\beta$), the fault dip ($f_\beta$), maximum burial depth ($z_\text{max}$) and faulting depth ($z_f$). Marginal probability distributions for another set of intermediate variables including fault thickness $f_T$, residual friction angle ($\phi_r^\prime$), critical shale smear factor (SSFc), porosity ($n$), permeability ($k$), and permeability anisotropy ratio ($k^\prime$), are generated and used to place fault materials and assign their properties. Finally, the permeability for the studied throw window is computed in the coarse grid by flow-based upscaling of the fine grid permeability using the MATLAB Reservoir Simulation Toolbox (MRST~\citep{lie2019introduction}). Repeated realizations of this process result in probability distributions for $k_{xx}$ and $k_{zz}$. We refer the readers to \citet{salo2023fault} and its supplemental material for details and notations.

\begin{figure}
\centering
\includegraphics[trim={15cm 0 15cm 0},clip, height = 0.35\textwidth]{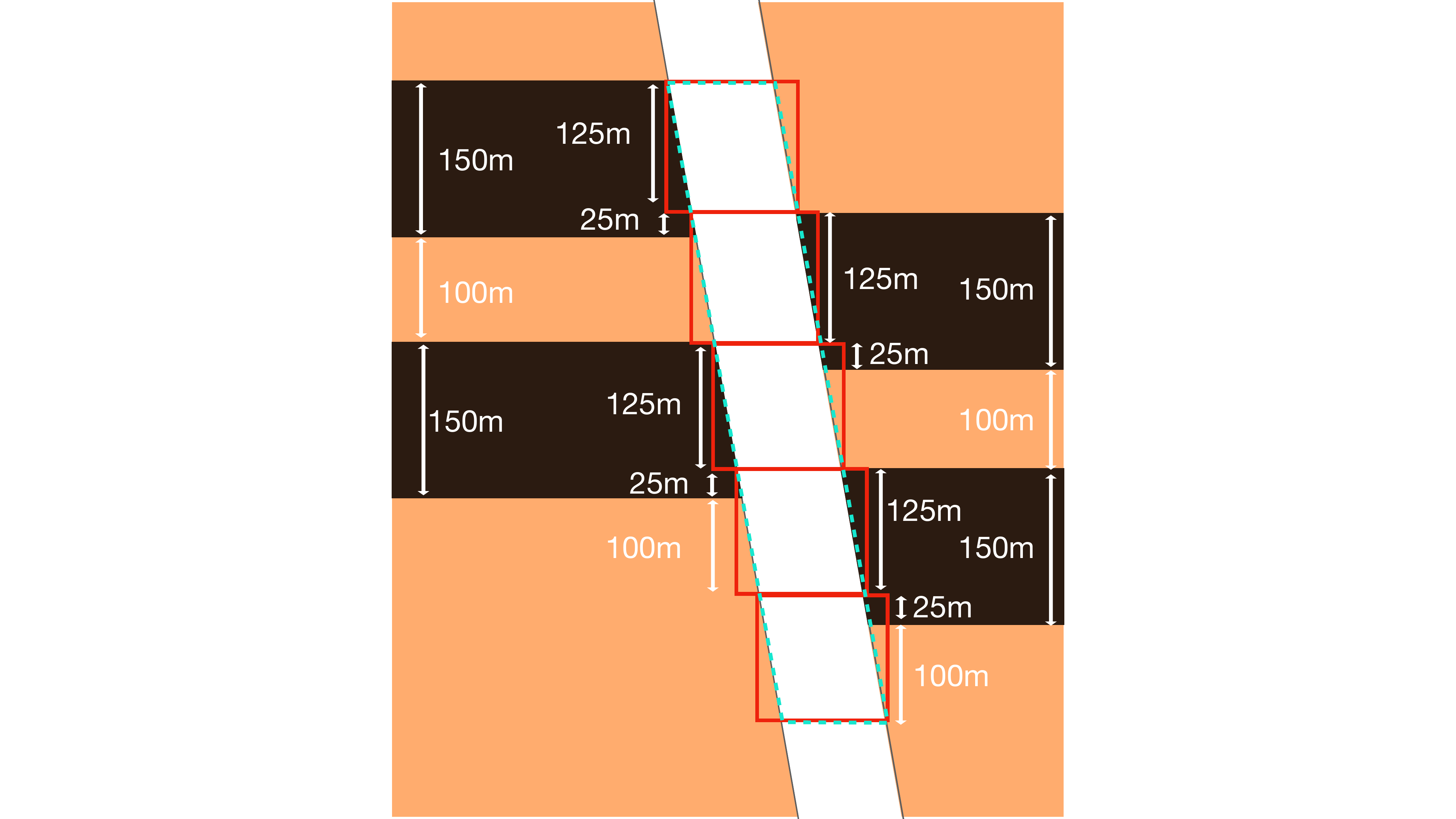}
\includegraphics[height = 0.35\textwidth]{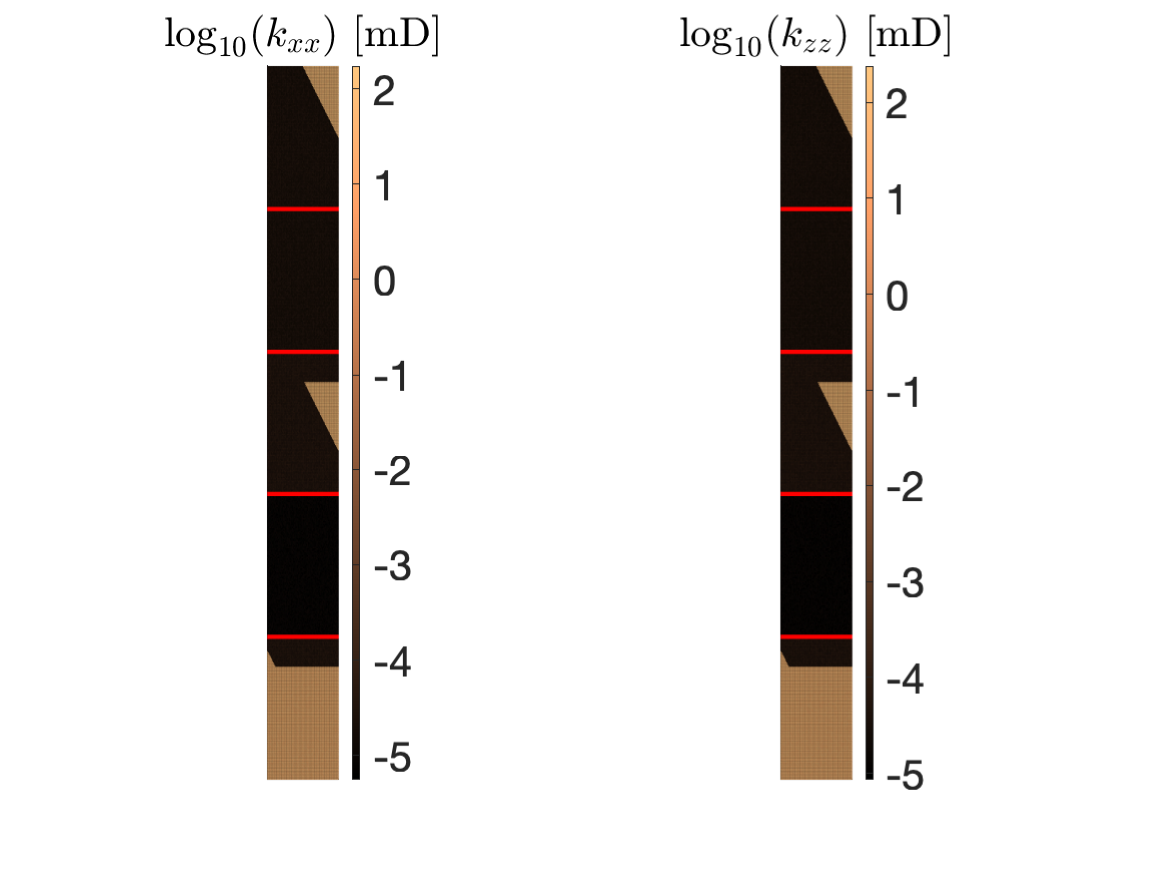}
\caption{Left: identification of 5 throw windows from the domain configuration (Figure~\ref{fig:domain}); Right: One example realization of material placement in the 5 windows (cyan dashed quadrilateral of the left figure). }
\label{fig:5window}
\end{figure}

\begin{figure}
\centering
\includegraphics[width = 0.8\textwidth]{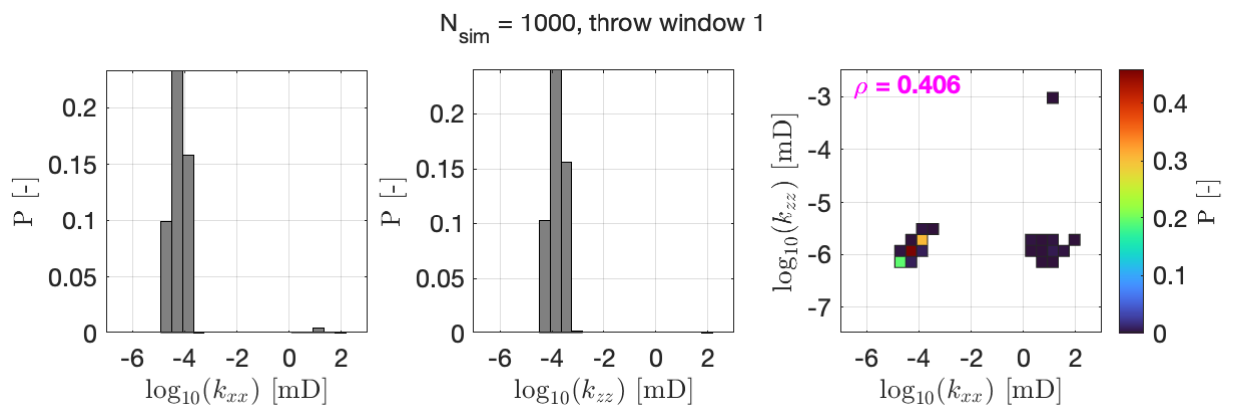}
\includegraphics[width = 0.8\textwidth]{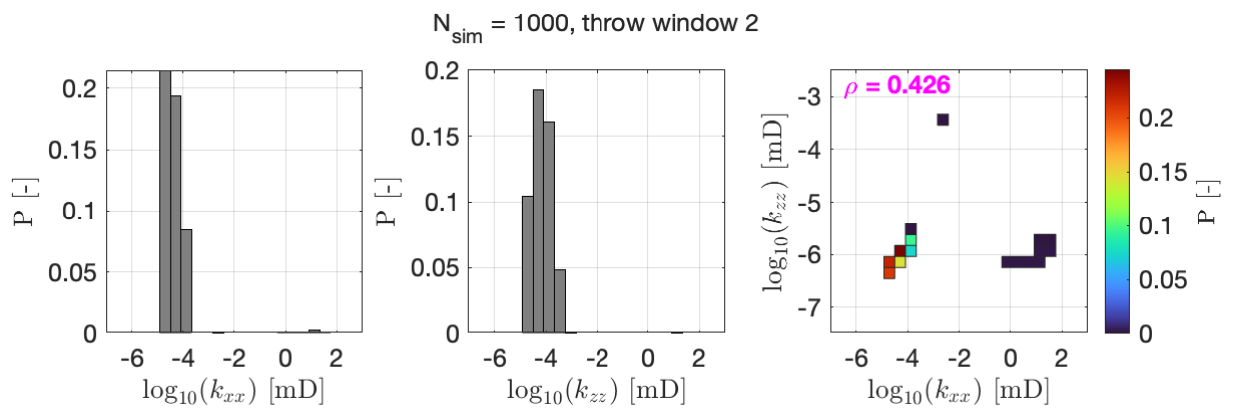}
\includegraphics[width = 0.8\textwidth]{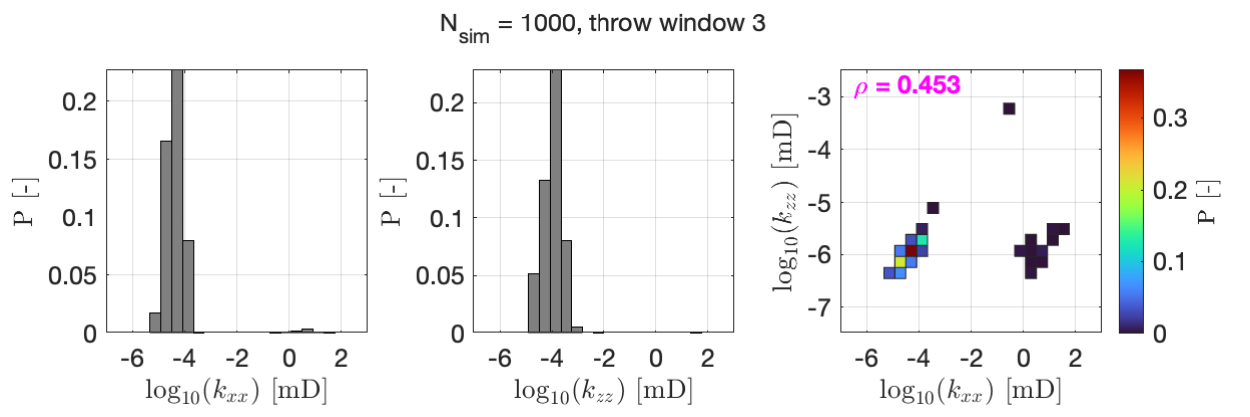}
\includegraphics[width = 0.8\textwidth]{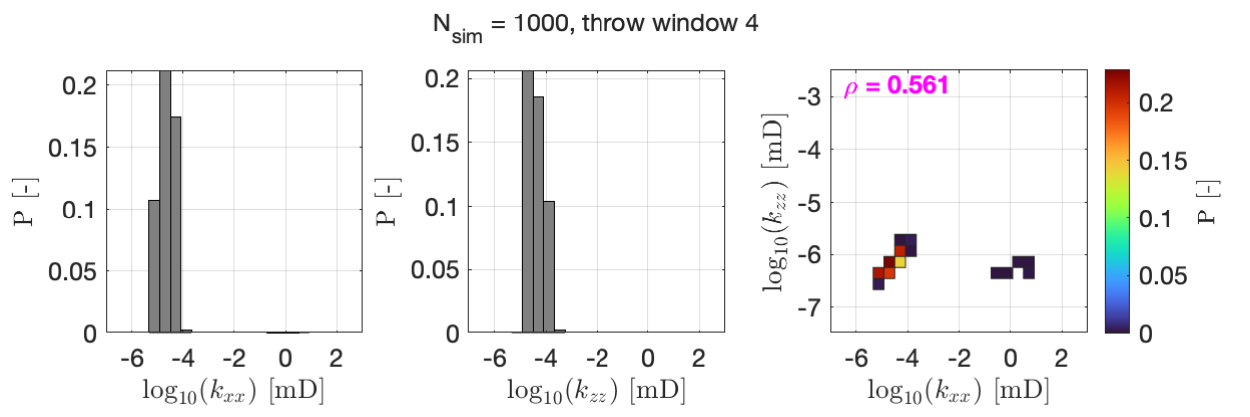}
\includegraphics[width = 0.8\textwidth]{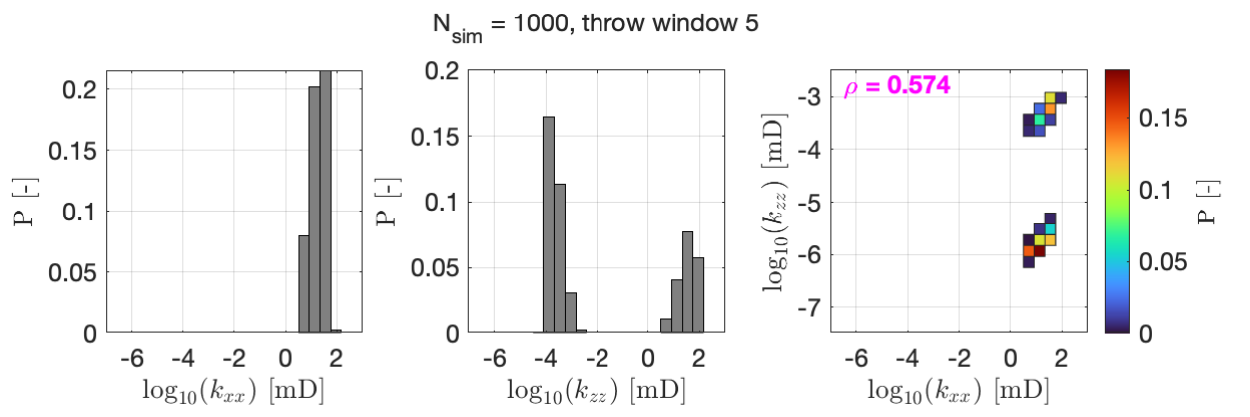}
\caption{Marginal probability distributions and pair-wise correlations of $\{(k_{w,xx},k_{w,zz})\}$ for all 5 fault windows, $w$ $=1,2,3,4,5$.}
\label{fig:kf_distribution}
\end{figure}

Given the configuration of the model in Figure~\ref{fig:domain}, five throw windows are identified (red boxes in Figure~\ref{fig:5window}). The input clay fraction of the caprock is set to be $0.4$ and the clay fraction of other aquifers is set to be $0.2$. The faulting depth $z_f = 200$ m in all windows. To ensure consistent material placement across throw windows, the value of fault thickness $f_T$ (intermediate variable) in one realization is the same for all 5 throw windows, making the $10$ values $\{(k_{w,xx},k_{w,zz})\}_{w=1}^5$ correlated. An example of the material placement in the 5 windows is shown in Figure~\ref{fig:5window}. Finally, repeated realizations of a set of 10 values $\{(k_{w,xx},k_{w,zz})\}_{w=1}^5$ are obtained by flow-based upscaling within each window~\citep{durlofsky1991numerical}. Figure~\ref{fig:kf_distribution} shows the distributions of the $10$ values from $N_\text{sim} = 1000$ samples. Our convergence test shows that fewer samples often result in an inaccurate representation of the probability distributions and the omission of extreme values.

\section{Data-driven Modeling of Quantities of Interest}\label{sec:3}
Let us consider the governing PDE~\eqref{eq:PDE} after spatial discretization on grid points $\{\boldsymbol \xi_i\}_{i=1}^{N_{\boldsymbol \xi}}$, where $\boldsymbol \xi_i\in \Omega$. We refer the readers to section~\ref{sec:MRST} for \ym{details on the design of our grid}. Denote by $\mathbf s(t;\mathbf p)\in \mathbb R^{N_{\mathbf s}}$ all state variables in~\eqref{eq:PDE} after spatial discretization. Then $\mathbf s$ satisfies an \ym{autonomous} system of ordinary differential equations (ODEs):
\begin{equation}\label{eq:ODE}
\frac{d\mathbf s}{dt} = \mathbf f(\mathbf s;\mathbf p),
\end{equation}
where the right-hand-side $\mathbf f:\mathbb R^{N_{\mathbf s}}\to \mathbb R^{N_{\mathbf s}}$ is determined by the governing PDE~\eqref{eq:PDE}, the choice of grid, and the simulator (i.e., the way to discretize spatial derivatives). Although $\mathbf f$ is known, it is an extremely high dimensional function ($N_{\mathbf s}\gg 1$), difficult to represent explicitly, and expensive to compute. The vector $\mathbf p = [\mathbf k_f,\lambda,k_s,\phi_s,c_r]\in \boldsymbol \Gamma\subset \mathbb R^{N_{\mathbf p}}$ represents the uncertain parameters, where $\mathbf k_f = \{(k_{w,xx},k_{w,zz})\}_{w=1}^5$ are generated from PREDICT as described in section~\ref{sec: uncertainties}. Hence we have $N_{\mathbf p} = 14$ uncertain parameters.

From the description in section~\ref{sec:risk}, we observe that the quantities that are needed for the study of CO$_2$ leakage and fault instability are only the state variables within the fault. In particular, we are interested in the following quantities of interest, $$\mathbf q(t;\mathbf p) =  \bigl (T_{s,\text{max}}(t;\mathbf p), \Delta p_\text{max}(t;\mathbf p), \Delta M_\text{b}(t;\mathbf p) \bigr ) \in \mathbb R^{N_\mathbf q},$$ with $N_{\mathbf q} = 3$. We describe these quantities as follows:
\begin{itemize}
\item $T_\text{max}(t;\mathbf p)$: the maximum value of the slip tendency defined in~\eqref{eq:slip-tendency} along the fault. The slip tendency increases with time due to injection and $T_\text{max}(t;\mathbf p)$ is reached at the bottom boundary of the aquifer ($1550$m) at each time instance. See Figure~\ref{fig:3-d} for example. $T_\text{max}(t;\mathbf p)$ needs to be monitored as a risk indicator of induced seismicity since slip will be induced once this value exceeds the coefficient of static friction $\mu_s$ (Figure~\ref{fig:Coulomb-diagram}).
\item $\Delta p_\text{max}(t;\mathbf p)$: the maximum value of the pressure buildup along the fault. The pressure buildup increases uniformly within the storage aquifer (between $1450$m and $1550$m) due to injection. See Figure~\ref{fig:3-a} for a representative example. Monitoring pressure buildup provides critical information for reservoir management and safe operations. It is required in regulatory compliance as well.
\item $\Delta M_\text{b}(t;\mathbf p)$: the total mass of brine leaking from the storage reservoir. Since the CO$_2$ plume migrates only a short distance away from the injection well and never reaches the fault during our considered time scale (Figure~\ref{fig:2-a}), we do not have concerns of CO$_2$ leakage. Instead, we monitor the brine leakage from the storage aquifer to assess the potential risk that fresh groundwater in other aquifers may be contaminated by the injection-formation brine.
\end{itemize}

In short-hand notation, our quantities of interest come from post-processing of the discrete state variables at the fault, i.e., 
\begin{equation}\label{eq:QoI}
\mathbf q (t;\mathbf p) = \mathcal M(\mathbf s_f(t;\mathbf p)),
\end{equation}
where $\mathcal M$ is the post-processing map (e.g., the computation of ~\eqref{eq:shear-normal2} and~\eqref{eq:slip-tendency}) and $\mathbf s_f$ represents the discretized states $\mathbf s$ within the fault domain $\Omega_f$. 

An exact mathematical model for the evolution of $\mathbf q(t;\mathbf p)$ usually does not exist. Our goal is to build a numerical model that approximates the evolution of $\mathbf q(t;\mathbf p)$ using a data-driven method. Once such numerical model is built, one can approximate the evolution of $\mathbf q(t;\mathbf p)$ directly without going through the expensive full-field computation of~\eqref{eq:ODE} and its mapping to the fault domain in~\eqref{eq:QoI}. The data-driven method that we choose to approximate the evolution of $\mathbf q(t;\mathbf p)$ is the FML method. Below, we review the methodology and adapt it to our problem.

\subsection{Flow Map Learning Methods}
Let us assume that time sequences of the QoIs are available at a finite collection of sampled parameter points $\{\mathbf p^{(1)},\ldots, \mathbf p^{(N_\text{budget})}\}$, in the following form,
\begin{equation}\label{eq:data}
\begin{aligned}
&\mathbf q_0(\mathbf p^{(1)}),\ldots,\mathbf q_{L_1}(\mathbf p^{(1)});\\
&\mathbf q_0(\mathbf p^{(2)}),\ldots,\mathbf q_{L_2}(\mathbf p^{(2)});\\
&\ldots \ldots \\
&\mathbf q_0(\mathbf p^{(N_\text{budget})}),\ldots,\mathbf q_{L_{N_\text{budget}}}(\mathbf p^{(N_\text{budget})}).\\
\end{aligned}
\end{equation}
Here we use the notation $\mathbf q_n(\mathbf p^{(i)}) \equiv \mathbf q(t_n;\mathbf p^{(i)})$. Each sequence represents an evolution trajectory data of $\mathbf q$ at a sampled parameter point $\mathbf p^{(i)}\in \boldsymbol \Gamma, i = 1,\ldots , N_\text{budget}$ with $\boldsymbol \Gamma$ being the parameter domain of interest. $N_\text{budget}\geq 1$ is the total number of such trajectories that one can afford to generate from the expensive simulation of~\eqref{eq:ODE} (e.g., the MRST simulator in our case) and post-processing~\eqref{eq:QoI}. For simplicity, we assume that the data are collected at time instances with a uniform time step $\Delta t$ for every trajectory, i.e., 
\begin{equation}
\Delta t \equiv t_n-t_{n-1},\quad \forall n.
\end{equation}
Note that each of the $i$-th trajectories, $i = 1,\ldots , N_\text{budget}$, has its own initial condition $\mathbf q(t_0; \mathbf p^{(i)})$ due to the dependence on $\mathbf p^{(i)}$. The \textit{length} of each trajectory $L_i$ also depends on $\mathbf p^{(i)}$, as the simulation is allowed to proceed only while $T_{s,\text{max}}(t;\mathbf p^{(i)} )< 1$.

The initial time $t_0$ is the same for each trajectory in our example and thus time variables are eliminated in the trajectory data~\eqref{eq:data}. However, this is not a necessity, meaning that each trajectory can be collected from different initial time. A key design principle of FML is that time variable shall not explicitly present itself in the process~\citep{churchill2023flow}. This implies that the absolute time $t$ is not present in the learning process and only the relative time shifts among different data entries matter.
This time variable removal design is the key to the success of FML in making long-term predictions and extrapolating in the time horizon.

Our goal is to construct a numerical time-marching model
\begin{equation}\label{eq:FML}
\mathbf q_{n+1}(\mathbf p) = \mathcal F(\mathbf q_n, \ldots ,\mathbf q_{n-n_M};\mathbf p),\quad n\geq n_M\geq0, \quad \forall \mathbf p\in \boldsymbol \Gamma,
\end{equation}
such that for any unsampled parameter $\mathbf p^*\in \boldsymbol \Gamma$ and given proper initial conditions on $\mathbf q(t_0;\mathbf p^*)$, the model prediction is an accurate approximation to the true solution at each time instances $t_0,\ldots , t_{L_*}$, where we want $L_{*}\gg 1$ to monitor the evolution of the QoIs until the critical point of slip. Here $n_M$ is an integer called the \textit{memory step} and $n_M \Delta t = T_M>0$ is the \textit{memory length} for the system. Accurate modeling of $\mathbf q$ would require a certain length of memory $T_M>0$ (i.e., $n_M>0$) because $\mathbf q$ is post-processed from $\mathbf s_f\subset \mathbf s$ according to~\eqref{eq:QoI}. This general FML formulation~\eqref{eq:FML} is motivated by the Mori-Zwanzig formalism~\citep{mori1965transport,zwanzig1973nonlinear} together with a parametric design, and the detailed derivation is available in~\citet{fu2020learning,qin2021deep}. The appropriate choice of $n_M$, or equivalently $T_M$, is problem dependent, which we investigate numerically for our problem of interest (see section~\ref{sec:results}). If the learning target is the evolution of the full set variable $\mathbf s$, then the FML model~\eqref{eq:FML} reduces to the special case of $n_M = 0$. However, learning $\mathbf s$ requires a model of much larger size ($N_{\mathbf s}\gg N_{\mathbf q}$) and post-processing will be needed afterwards to provide comprehensive information for risk analysis, making the modeling less direct and efficient.

 \subsection{Data Preparation}
 The mapping $\mathcal F: \mathbb R^{N_{\mathbf q}\times (n_M+1) +N_{\mathbf p}}\to \mathbb R^{N_{\mathbf q}}$ in~\eqref{eq:FML} is unknown. In order to learn $\mathcal F$, we choose trajectories of length 
 \begin{equation}
 n_\text{data} = n_M+2 +n_\text{MSL}
 \end{equation}
 from the raw dataset~\eqref{eq:data}, where $n_\text{MSL}\geq 0$ is needed for a  multi-step loss (see section~\ref{sec:loss}). Our training dataset thus consists of such ``trajectories'' of length $(n_M+2+n_\text{MSL})$,
 
 \begin{equation}\label{eq:data2}
\{\mathbf q_0^{(j_i)}(\mathbf p^{(i)}),\ldots ,\mathbf q_{n_M+1+n_\text{MSL}}^{(j_i)}(\mathbf p^{(i)}), \},\quad j_i = 1,\ldots , n_\text{burst},\quad i = 1,\ldots , N_\text{budget},
\end{equation}
where $\mathbf q_0^{(j_i)}(\mathbf p^{(i)})$ is any data entry in the raw dataset~\eqref{eq:data}, and $\mathbf q_1^{(j_i)}(\mathbf p^{(i)})$ to $\mathbf q_{n_M+1+n_\text{MSL}}^{(j_i)}(\mathbf p^{(i)})$ are the $(n_M+1+n_\text{MSL})$ entries immediately following. This requires that the length of each trajectory in the raw dataset~\eqref{eq:data} be sufficiently long, $L_i\geq n_\text{data}, \forall i$. If a trajectory in~\eqref{eq:data} has length $L_i>n_\text{data}$, then it is possible to choose multiple $n_\text{data}$-length trajectories into the training dataset~\eqref{eq:data2}. In this case, we can randomly choose, with a uniform distribution, $n_\text{burst} \geq 1$ segments of $n_\text{data}$ as consecutive entries to include in the training dataset~\eqref{eq:data2}. The total number of training trajectories in~\eqref{eq:data2} is fixed as $N = n_\text{burst}\cdot N_\text{budget}$. On one hand, increasing the value of $n_\text{burst}$ can boost the size of the training dataset~\eqref{eq:data2}. On the other hand, the $n_\text{burst}$ trajectories from the same trajectory in the raw dataset~\eqref{eq:data} represent a set of highly clustered entries in the domain. This situation is unfavorable for robust function approximation and thus a smaller $n_\text{burst}$ is preferred if the raw dataset~\eqref{eq:data} is easy to generate (i.e., if one can afford a large $N_\text{budget}$). For our problems in particular, we have a fixed budget for simulating~\eqref{eq:data} and would choose a value of $n_\text{burst}$ that can keep the size of training dataset~\eqref{eq:data2} $N$ reasonably large.

\subsection{Loss Function}\label{sec:loss}
First, let us consider a single-step loss function, where $n_\text{MSL} = 0$. Given the training dataset~\eqref{eq:data2}, the learning goal of finding $\mathcal F$ can be formulated as minimizing the mean-squared (MSE) loss
\begin{equation}\label{eq:ss-loss}
\min_{\Theta}\frac{1}{N_\text{budget}n_\text{burst}}\sum_{i = 1}^{N_\text{budget}}\sum_{j_i = 1}^{n_\text{burst}}\left\|\mathbf q_{n_M+1}^{(j_i)}(\mathbf p^{(i)})-\mathcal F_{\Theta}\left(\mathbf q_0^{(j_i)}(\mathbf p^{(i)}),\ldots ,\mathbf q_{n_M}^{(j_i)}(\mathbf p^{(i)})\right)\right\|^2,
\end{equation}
where $\Theta$ \ym{denotes the vector of parameters describing candidate functions $\mathcal F$.}

We generalize to a multi-step loss function to enhance the stability and accuracy of the FML model for long-term prediction. For $n_\text{MSL}>0$, one can conduct the system predictions by the FML model for $(n_\text{MSL}+1)$ steps:
\begin{equation}
\begin{aligned}
&\tilde{\mathbf q}_0^{(j_i)}(\mathbf p^{(i)}) = \mathbf q_0^{(j_i)}(\mathbf p^{(i)}),\ldots ,\tilde{\mathbf q}_{n_M}^{(j_i)}(\mathbf p^{(i)}) = \mathbf q_{n_M}^{(j_i)}(\mathbf p^{(i)}),\\
&\tilde{\mathbf q}_{n+1}^{(j_i)} = \mathcal F_\Theta\left(\tilde{\mathbf q}_{n-n_M}^{(j_i)},\ldots , \tilde{\mathbf q}_{n}^{(j_i)}\right),\quad n = n_M,\ldots ,n_M+n_\text{MSL},
\end{aligned}
\end{equation}  
The averaged loss against $\mathbf q_{n_M+1}^{(j_i)},\ldots , \mathbf q_{n_M+1+n_\text{MSL}}^{(j_i)}$ is then computed and~\eqref{eq:ss-loss} is generalized to
\begin{equation}\label{eq:ms-loss}
\min_{\Theta}\frac{1}{N_\text{budget}n_\text{burst}(n_\text{MSL}+1)}\sum_{i = 1}^{N_\text{budget}}\sum_{j_i = 1}^{n_\text{burst}}\sum_{k=0}^{n_\text{MSL}}\left\|\mathbf q_{n_M+1+k}^{(j_i)}(\mathbf p^{(i)})-\tilde{\mathbf q}_{n_M+1+k}^{(j_i)}(\mathbf p^{(i)})\right\|^2.
\end{equation}

\subsection{Deep Neural Networks}
\begin{figure}
\centering
\includegraphics[width = 0.8\textwidth]{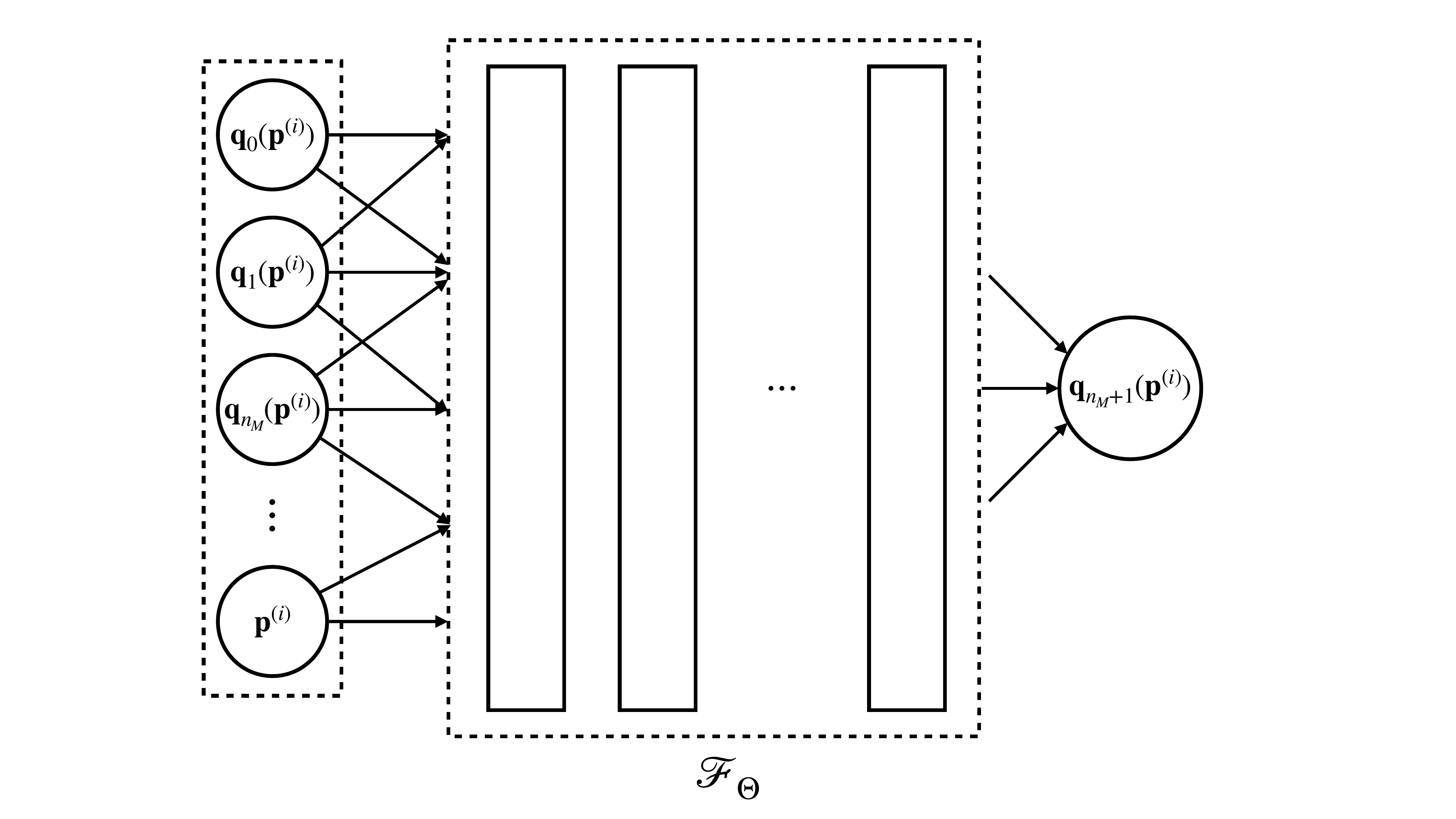}
\caption{Basic DNN structure for FML with memory and parameters.}
\label{fig:DNN}
\end{figure}

To numerically approximate $\mathcal F$, we choose $\mathcal F_\Theta$ to be a deep neural network (DNN), where $\Theta$ represents the parameters (i.e., weights and biases) of the network. The DNN training is thus conducted by solving~\eqref{eq:ms-loss} for the parameters $\Theta$. The DNN structure for FML is illustrated in Figure~\ref{fig:DNN}. There is no need for specialized neural network architectures and details on our choice of network architecture are given in section~\ref{sec:NN-arch}. 

The most effective way to enhance reproducibility of the FML models is ensemble averaging~\citep{churchill2023robust}. With everything else fixed, the initial random seeds for the optimizer can be varied and $N_\text{model}>1$ DNN models can be trained independently. During testing, each independent DNN model starts with the same initial conditions and evolves for one time step. The results are then averaged and provided to all models as the same new initial conditions in the next time step.

\section{Results and Discussion}\label{sec:4}
In this section, we present the results of our study with details in approaches to sampling the uncertain parameters and the configurations of both simulation tools and neural networks. We also thoroughly analyze the quantities of interest from uncertainty quantification, and discuss the implications in monitoring fluid leakage and fault instability and the robustness of NN in real time management of CO$_2$ storage projects.

\subsection{Parameter Sampling}
First, we assume that the uncertain parameters are  $\mathbf p = [\mathbf k_f,\lambda,k_s,\phi_s,c_r]\in \boldsymbol \Gamma \subset \mathbb R^{N_{\mathbf p}}$, where $\mathbf k_f = \{(k_{w,xx},k_{w,zz})\}_{w=1}^5$ follow the distribution represented by PREDICT (section~\ref{sec: uncertainties}) and $\lambda \sim \mathcal U[0.6, 0.8]$, $\log_{10}k_s \sim \mathcal U [ -14+\log_{10}5,-13+\log_{10}5]$, $\phi_s\sim \mathcal U[0.1,0.3]$, $\log_{10}c_r\in \mathcal U[-5+\log_{10}3, -4+\log_{10}3]$. Here $\mathcal U[a,b]$ stands for uniform distribution between $a$ and $b$ and we assume that $\mathbf k_f, \lambda, k_s, \phi_s, c_r$ are independent from each other, while $\{(k_{w,xx},k_{w,zz})\}$ are dependent and correlated. $N_\text{sim} = 1000$ samples $\{\mathbf p^{(i)}\}_{i=1}^{N_\text{sim}}$ are randomly drawn from $\boldsymbol \Gamma$ and used as representative of the input uncertainties. Due to the complex distribution of $\mathbf k_f$ and the correlations between $\{(k_{w,xx},k_{w,zz})\}$, a space-filling design modified from~\citet{johnson1990minimax} is used to select, \ym{from this larger candidate set,} the $N_\text{budget} = 100$ samples used to produce the raw training dataset~\eqref{eq:data}. We will show that the surrogate model learned from the training \ym{dataset produced by this space-filling design}  has better prediction performance than the model learned from a random-design training dataset of the same size. We also note that the space-filling design is constructed \textit{prior} to running any expensive model simulations.

\begin{algorithm}
\caption{Modified space-filling design}\label{alg:space-filling}
Input: $\{\mathbf p^{(i)}\}_{i=1}^{N_\text{sim}}$, $\mathcal D = \emptyset$, $\mathcal I =  \{ 1,\ldots , N_\text{sim}\}$
\begin{itemize}
\item $i_1 = \arg\max_{i \in \mathcal I}\sum_{j = 1}^{N_\text{sim}}\|\mathbf p^{(i)}-\mathbf p^{(j)}\|_2$, $\mathcal D = \{\mathbf p^{(i_1)}\}$, $\mathcal I = \mathcal I \setminus \{i_1\}$.
\item For $j = 2,\ldots , N_\text{budget}$
\begin{itemize}
\item $i_j = \arg\max_{i\in \mathcal I}\left(\min_{\mathbf p^{(j)}\in \mathcal D} \|\mathbf p^{(i)}-\mathbf p^{(j)}\|_2\right),$
\item $\mathcal D = \mathcal D \cup \{\mathbf p^{(i_j)}\}$, $\mathcal I = \mathcal I\setminus \{i_j\}$.
\end{itemize}

\end{itemize}
Output: $\mathcal D$.
\end{algorithm}

\subsection{MRST Simulation}\label{sec:MRST}
\begin{figure}
\centering
\includegraphics[width = 0.45\textwidth]{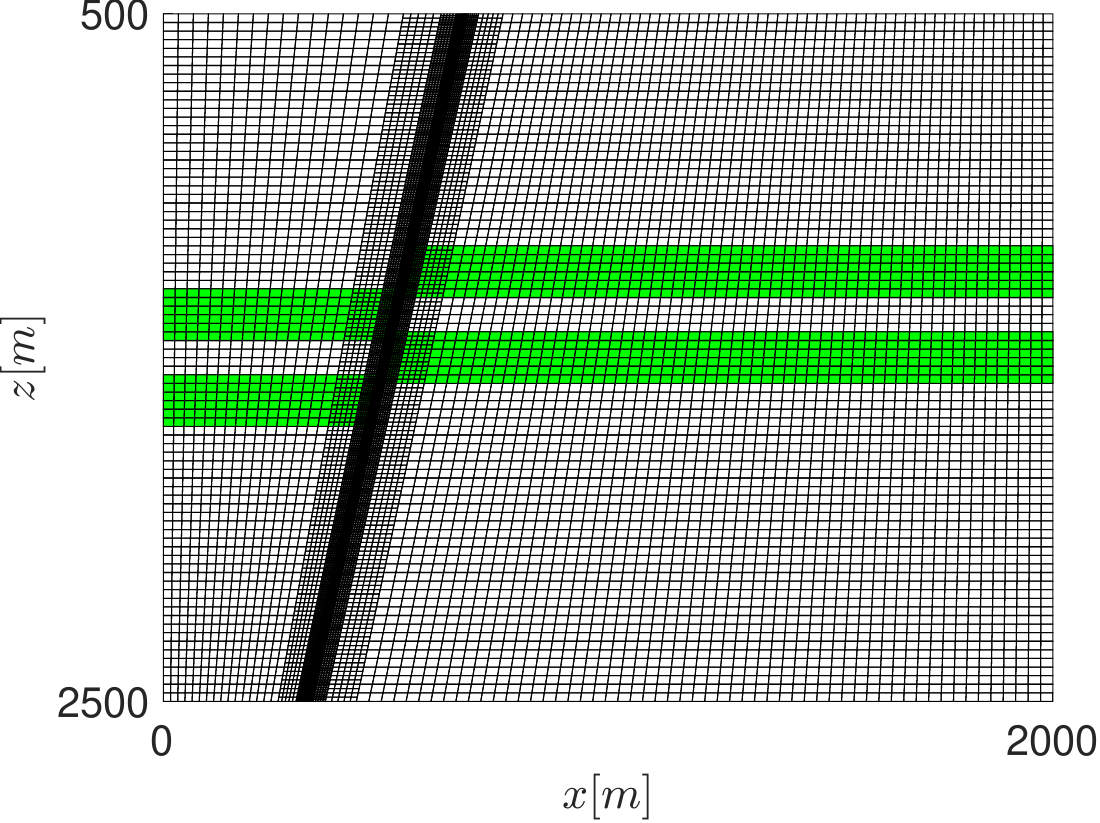}
\includegraphics[width = 0.45\textwidth]{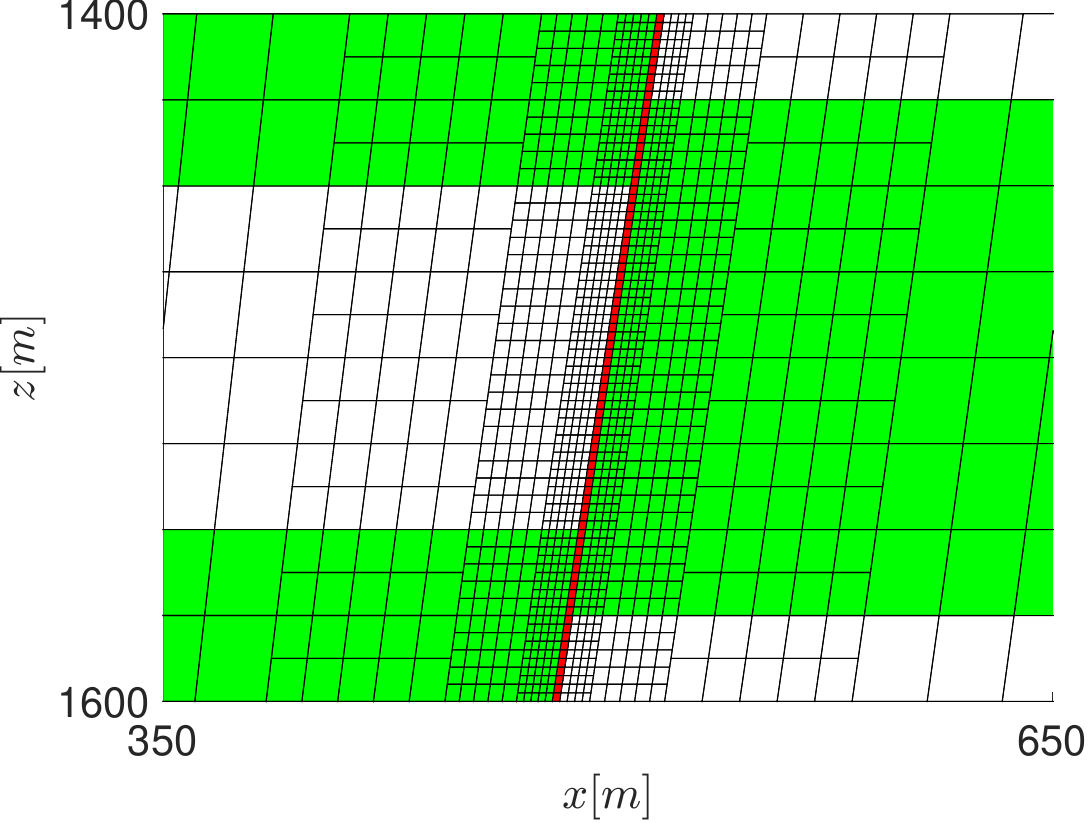}
\caption{Visualization of the computational grid for the entire domain (left) and a zoomed view near the fault (right). Green: caprocks; Red: fault.}
\label{fig:grid}
\end{figure}

The training data and the reference solution of the test data are generated from MRST simulations. First, we discretize the domain Figure~\ref{fig:domain} in multiblock grids in which different types of structured subgrids are glued together (Figure~\ref{fig:grid}). We start by generating four different block types using \textit{cartGrid}: the finest grid within and near the fault (with gridblocks of size $2.5$ m $\times$ $2.5$ m), the coarsest grid far away from the fault (with gridblocks of size $25$ m $\times$ $25$ m), and the middle-size grids bridging the finest and the coarsest (with gridblocks of size $5$ m $\times$ $5$ m and $12.5$ m $\times$ $12.5$ m). This design provides a local refinement near the fault, which is critical to a stable and accurate solution of the multiphase flow equations. Once these blocks are generated, we glue them together using \textit{glue2DGrid} and then apply a ``skew'' function to the coordinates so that the grid is aligned with the fault dip angle in $\Omega_f$.

The two-way coupled fluid flow and geomechanics equations are then modeled in MRST using the \textit{ad-mechanics} module~\citep{lie2021advanced}. One can choose fully coupled or operator splitting (fixed-stress splitting scheme) to implement time evolution of the coupled system. In our simulations, we use fully coupled scheme but observe no significant difference between the two schemes. The mechanics equations are discretized using first-order virtual element method (VEM), whereas the flow equations are discretized using standard finite volumes with two-point flux and a fully implicit first-order Euler scheme in time. The three-phase black-oil model is used in the flow problem where CO$_2$ is treated as ``dry gas'' and brine is treated as ``live oil'', therefore capable of modeling CO2 dissolution in brine~\citep{lie2019introduction}.

\begin{figure}
\centering
\includegraphics[width = 0.45\textwidth]{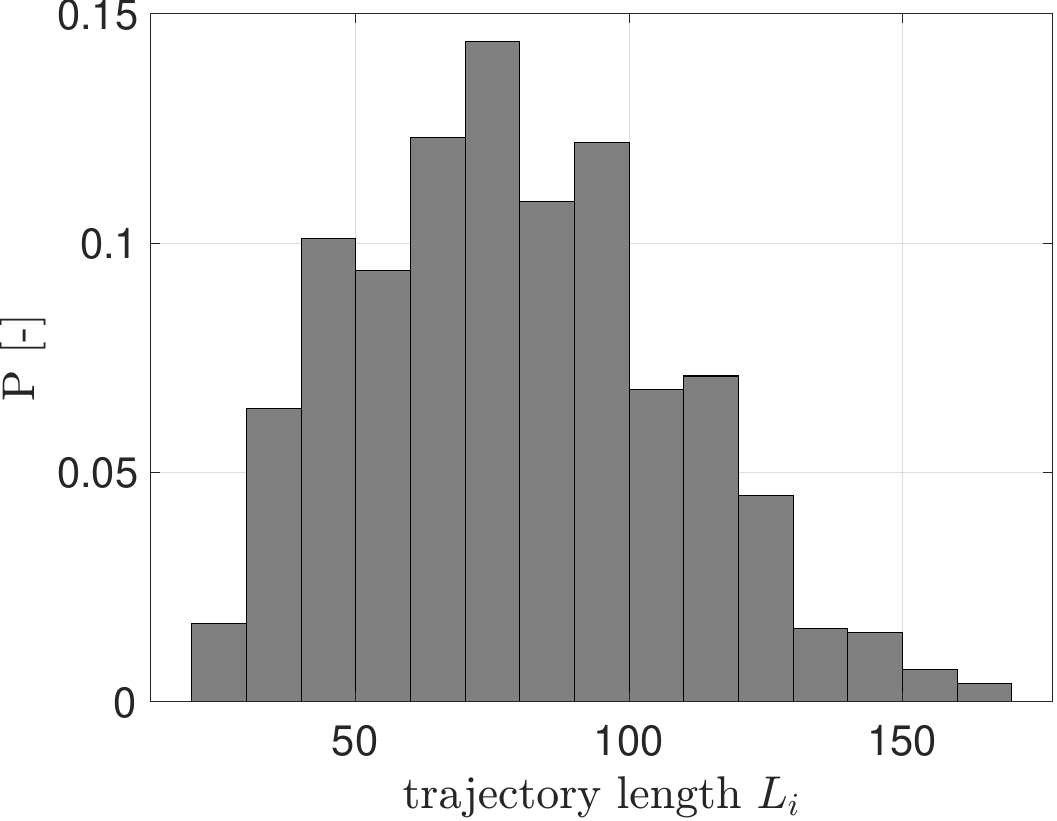}
\caption{Distribution of trajectory lengths $\{L_i\}_{i=1}^{N_\text{sim}}$ determined by the criteria $T_\text{max}(t;\mathbf p^{(i)})<0.6$ in the MRST simulations.}
\label{fig:L06}
\end{figure}

Along each trajectory with input $\mathbf p^{(i)}$, the simulation is stopped when $T_{s,\text{max}}(t;\mathbf p^{(i)}) = 0.6$ (exceeding the standard value of the friction coefficient $\mu_s$). The outputs are collected every $4$ hours ($\Delta t \equiv 4$ hours) after CO$_2$ injection starts. Figure~\ref{fig:L06} shows a histogram of the trajectory lengths $\{L_i\}_{i=1}^{N_\text{sim}}$, reflecting a wide range of the time horizon in our dataset. We emphasize that these trajectory lengths cannot be determined \emph{a prior}. It is therefore crucial to construct a time-marching surrogate model as in~\eqref{eq:FML} instead of one with a fixed time horizon. The time marching design avoids unphysical and unnecessary predictions beyond the critical point when the physics model is no longer valid.

Finally, the training dataset is collected from MRST simulations and post-processed in the form of~\eqref{eq:data} at the $N_\text{budget}$ parameter points $\mathcal D$ selected from Algorithm~\ref{alg:space-filling}. We also obtain the reference solution of the test dataset $\{\mathbf q(t;\mathbf p^{(i)})\}_{i=1}^{N_\text{sim}}$ from MRST simulation and post-processing, to evaluate the performance of our DNN surrogate model.

\subsection{Neural Network Hyperparameters}\label{sec:NN-arch}
We employ simple feed-forward fully connected neural network structures with $3$ layers and $10$ neurons per layer. In our experiments, we observe improved prediction performance with more memory steps; however, the overall efficiency—considering both the accuracy across the entire trajectory and the computational costs of the initial steps—may decline if the number of memory steps becomes too large. Therefore to balance the accuracy and efficiency, we decide to use $n_M = 20$ memory steps and $n_\text{MSL} = 10$ for multi-step loss. We have $N_\text{budget} = 100$ and $n_\text{burst} = 20$, making the total number of trajectories $N = 2000$. $N_\text{model} = 10$ DNN models are trained independently with various initial random seeds for the optimizer and the prediction results are averaged over all model predictions in the testing procedure. All data are mapped via logarithm function and scaled within $[-0.5, 0.5]$. This scaling is a critical pre-processing step in using DNN models as a large spread of values may result in large error gradient values causing weight values to change dramatically, making the learning process unstable~\citep{bishop1995neural}. We observe better performance when training the QoIs $T_{s,\text{max}}$, $\Delta p_\text{max}$ and $\Delta M_b$ separately than when training them together. Therefore, each DNN model for each QoI has a total number of $581$ trainable hyperparameters, which is much less than the number of training data trajectories. Each model is trained for $10\,000$ epochs with a learning rate of $10^{-4}$, and the best model based on training loss is saved.

\subsection{Uncertainty Analysis for the Quantities of Interest}\label{sec:results}
\begin{figure}
\centering
\includegraphics[width = \textwidth]{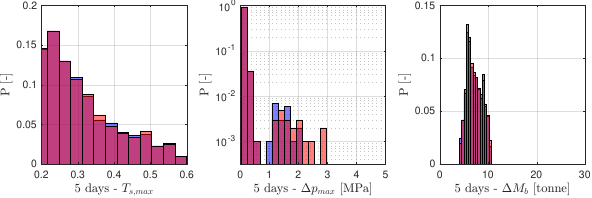}
\includegraphics[width = \textwidth]{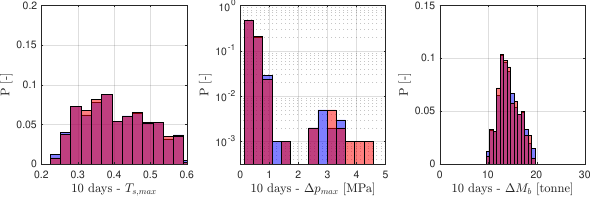}
\includegraphics[width = \textwidth]{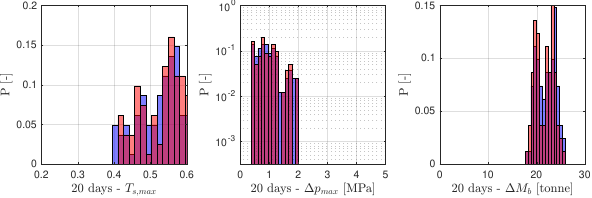}
\caption{Histograms of the different QoIs at $t = 5$ days, $10$ days and $20$ days. The probability distributions of the reference solutions are shown in blue and the counterpart of the FML predictions are shown in red.}
\label{fig:histogram}
\end{figure}

\begin{figure}
\centering
\includegraphics[width = \textwidth]{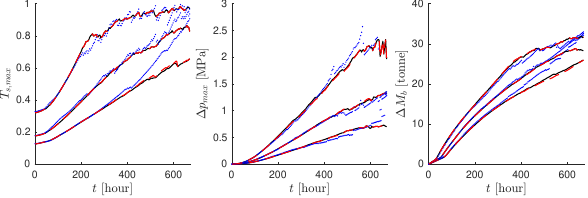}
\caption{Ensemble statistics from reference MRST simulations (black solid curves), FML predictions (red dashed curves) and $N_\text{budget}=100$ training data (blue dot curves). The lower, middle, and upper curves correspond to the $10$th, $50$th, and $90$th percentiles, respectively.}
\label{fig:percentile}
\end{figure}

Figure~\ref{fig:histogram} shows the comparison of the probability distributions obtained from reference full-physics Monte Carlo solutions and FML surrogate Monte Carlo solutions with the same number of ensembles ($N_\text{sim} = 1000$). It is remarkable that the surrogate model is able to capture the shape of the distributions of all the QoIs at all times, even though these distributions change significantly over time and exhibit multi-modal behavior. The difference between the two increases with time as errors (i.e, surrogate biases) accumulate over time steps. We plot the 10th, 50th and 90th percentiles of the QoIs in Figure~\ref{fig:percentile}. 
\ym{Estimating these percentiles \textit{directly} from the $N_\text{budget} = 100$ training dataset (blue dotted curves) provides a poor approximation to their true values, as represented by percentiles of the reference dataset (black solid curves).} \ym{Using the same $N_\text{budget}$ dataset to build a deep learning surrogate of the flow map, however, } can \ym{produce accurate ensembles of trajectories and thus approximate these statistics} with much higher accuracy (shown in red dashed curves). The computational cost of building this surrogate (i.e., generating $N_\text{budget}$ high-fidelity data and training the NN) is about one order of magnitude lower than that of the reference (i.e., generating $N_\text{sim}$ high-fidelity data).

To quantitatively assess the accuracy of each modeling approach, we define the relative $L^2$ error of each QoI at any test parameter $\mathbf p^{(i)}\in \{\mathbf p^{(i)}\}_{i=1}^{N_\text{sim}}$ as
\begin{subequations}
\begin{align}
&\varepsilon^{(i)}_{T_{s,\text{max}}} =\frac{ \sum_{n=1}^{L_i}\left(T_{s,\text{max}}(t_n;\mathbf p^{(i)})-\tilde{T}_{s,\text{max}}(t_n;\mathbf p^{(i)})\right)^2}{\sum_{n=1}^{L_i}\left(T_\text{max}(t_n;\mathbf p^{(i)}))\right)^2},\\
&\varepsilon^{(i)}_{\Delta p_\text{max}} =\frac{ \sum_{n=1}^{L_i}\left(\Delta p_\text{max}(t_n;\mathbf p^{(i)})-\Delta \tilde {p}_\text{max}(t_n;\mathbf p^{(i)})\right)^2}{\sum_{n=1}^{L_i}\left(\Delta p_\text{max}(t_n;\mathbf p^{(i)}))\right)^2},\\
&\varepsilon^{(i)}_{\Delta M_\text{b}} =\frac{ \sum_{n=1}^{L_i}\left(\Delta M_\text{b}(t_n;\mathbf p^{(i)})-\Delta \tilde{M}_\text{b}(t_n;\mathbf p^{(i)})\right)^2}{\sum_{n=1}^{L_i}\left(\Delta M_\text{b}(t_n;\mathbf p^{(i)}))\right)^2}.
\end{align}
\end{subequations}
The average errors $\bar \varepsilon_{T_{s,\text{max}}}$, $\bar \varepsilon_{\Delta p_\text{max}}$ and $\bar \varepsilon_{\Delta M_b}$ of the test dataset over the $N_\text{sim}$ trajectories are reported in Table~\ref{table:3}. With the space-filling design for training parameter values and hence training data trajectories, the  FML surrogate model yields accurate predictions for all test data points, even for relatively long simulation time. The surrogate model trained on randomly selected parameters, on the other hand, has less satisfactory performance and produces unstable predictions for some  trajectories, especially for $\Delta p_\text{max}$. In fact, there are a few test samples that have significantly larger $\varepsilon_{\Delta p_\text{max}}^{(i)}$ than others, making the overall average prediction error large. From Figure~\ref{fig:histogram}, we can observe that the probability distribution of $\Delta p_{max}$ exhibits two modes, and that both modes shift towards to larger values as injection continues. The second mode disappears in 20 days because the values reach the threshold, making the simulations stop before 20 days. The randomly sampled training data is unable to accurately identify the bifurcation in the parameter space that contributes to the two modes, making the FML surrogate model extremely inaccurate around the bifurcation region, as the outputs can be sent to the wrong cluster. In contrast, the distributions of $T_{s,\text{max}}$ and $\Delta M_b$ exhibit only one mode and are thus less affected by sub-optimal sampling. This highlights the importance of a good sampling strategy in $\boldsymbol \Gamma$ and a representative selection of the training dataset so that the neural network can generalize better, especially for challenging multi-modal distributions.

\begin{table}
\caption{Average prediction error of (i) FML model trained on randomly selected $N_\text{budget} = 100$ training trajectories; (ii) FML model trained on space-filling selection of the $N_\text{budget} = 100$ training trajectories.}
\begin{center}
\begin{tabular}{ l  c c c  } 
 \hline
 &$\bar \varepsilon_{T_{s,\text{max}}}$&$\bar \varepsilon_{\Delta p_\text{max}}$&$\bar \varepsilon_{\Delta M_b}$\\ 
   \hline
 random&$2.91\times 10^{-4}$&$27.88$&$4.55\times 10^{-4}$\\
  \hline
space-filling&$1.67\times 10^{-4}$&$2.20\times 10^{-3}$&$3.51\times 10^{-4}$\\
 \hline
\end{tabular}
\end{center}
\label{table:3}
\end{table}

Next, we use the FML surrogate for Monte Carlo estimation of variance-based global sensitivity analysis, which plays an important role in comprehensively characterizing how the uncertainty in the input parameters $\mathbf{p}$ propagates to the uncertainty in the QoIs $\mathbf{q}$. We use Sobol' variance-based sensitivity indices~\citep{sobol1993sensitivity}, which attribute portions of output variance not only to the influence of individual inputs but also to their interactions. Typically, the \textit{main effect} sensitivity indices and the \textit{total effect} sensitivity indices are estimated via Monte Carlo simulation using ``fixing methods''~\citep{saltelli1993sensitivity}. In addition to $\{\mathbf p^{(i)}\}_{i=1}^{N_\text{sim}}$, these methods use a second set of $N_\text{sim}$ independent realizations of $\mathbf p$ denoted by $\{\hat{\mathbf p}^{(i)}\}_{i=1}^{N_\text{sim}}$, where $\hat{\mathbf p}^{(i)}= (\hat{\mathbf k}_f^{(i)},\hat\lambda^{(i)},\hat k_s^{(i)},\hat \phi_s^{(i)},\hat c_r^{(i)})$. We define the n-tuples 
\begin{equation}
\begin{aligned}
      &\mathbf y_1^{(i)} = (\mathbf k_f^{(i)},\hat{\lambda}^{(i)},\hat k_s^{(i)},\hat \phi_s^{(i)},\hat c_r^{(i)}),\\
      &\mathbf y_2^{(i)} = (\hat{\mathbf k}_f^{(i)},\lambda^{(i)},\hat k_s^{(i)},\hat \phi_s^{(i)},\hat c_r^{(i)}),\\
      &\mathbf y_3^{(i)} = (\hat{\mathbf k}_f^{(i)},\hat{\lambda}^{(i)},k_s^{(i)},\hat \phi_s^{(i)},\hat c_r^{(i)}),\\
      &\mathbf y_4^{(i)} = (\hat{\mathbf k}_f^{(i)},\hat{\lambda}^{(i)},\hat k_s^{(i)},\phi_s^{(i)},\hat c_r^{(i)}),\\
      &\mathbf y_5^{(i)} = (\hat{\mathbf k}_f^{(i)},\hat{\lambda}^{(i)},\hat k_s^{(i)},\hat \phi_s^{(i)},c_r^{(i)}).
\end{aligned}
\end{equation}
Since $\mathbf k_f = \{(k_{w,xx},k_{w,zz})\}_{w=1}^5$ are generated from PREDICT respecting their mutual correlation, the sensitivity is evaluated by grouping them together.

The sample mean and variance for each QoI at $t_n$ are given by
\begin{equation}
    \mathbf E(t_n) = \frac{1}{N_\text{sim}}\sum_{i=1}^{N_\text{sim}}\mathbf q(t_n,\mathbf p^{(i)}), \quad \mathbf V(t_n) = \frac{1}{N_\text{sim}-1}\sum_{i=1}^{N_\text{sim}}(\mathbf q(t_n,\mathbf p^{(i)})-\mathbf E(t_n))^2,
\end{equation}
where $\mathbf E(t_n) = (E^{T_{s,\text{max}}}(t_n),E^{\Delta p_\text{max}}(t_n),E^{\Delta M_b}(t_n))^\top$ and $\mathbf V(t_n) = (V^{T_{s,\text{max}}}(t_n),V^{\Delta p_\text{max}}(t_n),V^{\Delta M_b}(t_n))^\top$.

We use a bias-corrected estimation of the Sobol' indices introduced by Owen~\citep{owen2013variance}
\begin{equation}
    s_j^{T_{s,\text{max}}}(t_n) = \frac{V_j^{T_{s,\text{max}}}(t_n)}{V^{T_{s,\text{max}}}(t_n)},\quad S_j^{T_{s,\text{max}}}(t_n) = \frac{T_j^{T_{s,\text{max}}}(t_n)}{V^{T_{s,\text{max}}}(t_n)}, \quad j = 1,\ldots , 5,
\end{equation}
with
\begin{equation}
\begin{aligned}
& V_j^{T_{s,\text{max}}} = \frac{2N_\text{sim}}{2N_\text{sim}-1}\left(\frac{1}{N_\text{sim}}\sum_{i = 1}^{N_\text{sim}}T_{s,\text{max}}(\mathbf p^{(i)})T_{s,\text{max}}(\mathbf y_j^{(i)}) -\left(\frac{E^{T_{s,\text{max}}}+\hat E^{T_{s,\text{max}}}}{2}\right)^2 +\frac{V^{T_{s,\text{max}}}+\hat V^{T_{s,\text{max}}}}{4N_\text{sim}}  \right)\\
& T_j^{T_{s,\text{max}}} = \frac{1}{2N_\text{sim}}\sum_{i=1}^{N_\text{sim}}\left(T_{s,\text{max}}(\hat{\mathbf p}^{(i)})-T_{s,\text{max}}(\mathbf y_j^{(i)})\right)^2
\end{aligned}
\end{equation}
where $E^{T_{s,\text{max}}}, \hat E^{T_{s,\text{max}}}$ and $V^{T_{s,\text{max}}}, \hat V^{T_{s,\text{max}}}$ are the sample means and variances, respectively, estimated using $\{\mathbf p^{(i)}\}_{i=1}^{N_\text{sim}}$ and $\{\hat{\mathbf p}^{(i)}\}_{i=1}^{N_\text{sim}}$. $s_j^{T_{s,\text{max}}}$ and $S_j^{T_{s,\text{max}}}$ are the Sobol' main effect sensitivity index and the Sobol' total effect sensitivity index of $T_{s,\text{max}}$ for input $j$, respectively. The Sobol' indices of $\Delta p_\text{max}$ and $\Delta M_b$ are evaluated analogously. We refer the readers to~\citet{dellino2015uncertainty} for a comprehensive review of global sensitivity analysis.

\begin{figure}
\centering
\includegraphics[width = \textwidth]{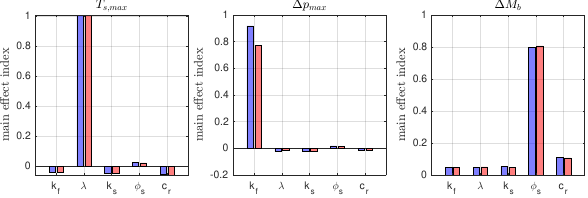}
\includegraphics[width = \textwidth]{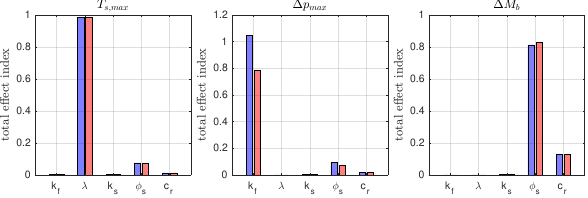}
\caption{Top: the Sobol’ main effect sensitivity indices of each QoI for each parameter at $t = 5$ days. Bottom: the Sobol’ total effect sensitivity indices of each QoI for each parameter at $t = 5$ days. The index values of the MC reference estimation are shown in blue and the counterparts of the FML surrogate estimation are shown in red with $N_\text{sim} = 1,000$.}
\label{fig:sensitivity}
\end{figure}

Monte Carlo estimation of Sobol' indices requires $(N_p^\prime+2)N_\text{sim}$ evaluations of the parameter-to-QoI map, where $N_p^\prime$ is the number of independent input parameters ($N_p^\prime = 5$ in our case). The computation becomes prohibitively expensive using the high fidelity simulation~\eqref{eq:ODE} and post-processing~\eqref{eq:QoI}. In fact, using the FML surrogate model for Monte Carlo estimation of Sobol' indices in this setting becomes essential. 

In  Figure~\ref{fig:sensitivity}, we compare the results of global sensitivity analysis with the full simulator and with the FML surrogate, fixing $N_\text{sim} =1000$ in both cases. This sample size is near the upper limit of what is computationally feasible with the full simulator.
Both estimation approaches show consistent results: the parameter $\lambda$ has a dominant effect on $T_{s,\text{max}}$ as it defines the initial ratio of the vertical stress and the horizontal stress; $\Delta p_\text{max}$ is highly sensitive to the values of $k_f$, which can be explained by the pressure equation; and the total mass of brine leakage $\Delta M_b$ is mostly responsive to the variations in the storage aquifer porosity $\phi_s$, as this parameter determines the pore volume. We can also observe secondary effects of porosity $\phi_s$ on $T_{s,\text{max}}$ and $\Delta p_\text{max}$ and secondary effects of rock compressibility $c_r$ on $\Delta M_b$. Some estimates of the small-magnitude Sobol' indices take negative values, however, indicating that $N_\text{sim} = 1000$ may be insufficient for accurate MC estimation.

\begin{figure}
\centering
\includegraphics[width = \textwidth]{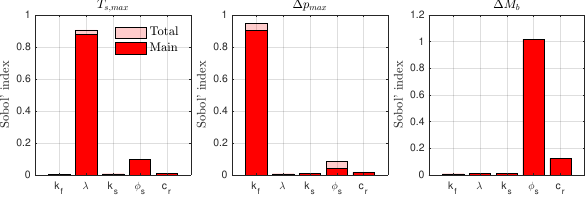}
\caption{Sobol’ sensitivity indices of each QoI for each parameter at $t = 5$ days, computed from $N_\text{sim} = 10,000$ MC estimation using the FML surrogate model.}
\label{fig:sensitivity2}
\end{figure}

To correct these negative estimates, we repeat the process using a larger number $N_\text{sim} = 10000$ of input parameters, using the trained FML surrogate model to efficiently compute the output QoIs with the requisite $(N_p^\prime + 2) N_{\text{sim}} = 70000$ runs. Figure~\ref{fig:sensitivity2} shows that a sufficiently large number of samples can correct the small negative values of the estimated Sobol' indices, with the accuracy of the FML surrogate guaranteeing consistent results with the full-physics simulation but at much lower computational cost. Here we do not show the results from full-physics simulation due to the prohibitive cost of simulating $70000$ such trajectories. More accurate estimation of the main- and total-effect Sobol' indices now also lets us better distinguish the influence of individual inputs from their interactions with each other; for instance, $\lambda$ individually \textit{and} its interactions with other parameters contribute to the variability of $T_{s,\text{max}}$.

\section{Conclusion}
In this work, we have developed a deep-learning-based surrogate model directly for the quantities of interest in a complex coupled model of multiphase flow and geomechanics during CO$_2$ storage. The surrogate model provides accurate, low-cost, long-term predictions for scenarios not seen in the training dataset. By incorporating past memory terms, the FML method is capable of directly learning unknown dynamics of the QoIs without considering the high dimensional full field. The FML surrogate, as a time-marching model, can provide a flexible prediction horizon and avoid unphysical predictions beyond the critical point when the physics model is no longer valid. With recent developments in stochastic modeling of properties in fault zones, investigating the effects of uncertainties from the fault zone and reservoirs can be significantly accelerated by this cheap but reliable surrogate model. With a space-filling design over the space of uncertain parameters, we can construct a small but informative training dataset and train a robust dynamical surrogate model over the parameter space. The trained surrogate model then produces a large number of accurate predictions at very low computational cost, improves estimates of ensemble statistics significantly over what would be feasible with the training dataset alone, and enables a sensitivity analysis that would otherwise be prohibitively expensive.

Our study provides probabilistic assessments of fault instability and fluid leakage, which can help guide decision-making and real-time management of CO$_2$ storage projects. Besides their use in forward uncertainty quantification, future applications of surrogate models include history matching and optimal control during CO$_2$ injection. Field data can be collected from monitoring wells and used in inference to reduced the uncertainties of the geological models and fault properties. Optimal injection plans (e.g., injection rates and well locations), as well as remedial actions in the event of unintended consequences, can be efficiently developed with reliable surrogate models.

\bibliography{CCS-UQ}

\end{document}